\documentclass[useAMS,usenatbib, a4]{mn2e}

\usepackage{graphicx}

\title[CO observations of star-forming galaxies in an overdensity at $z=1.5$]{Deep observations of CO line emission from star-forming galaxies in a cluster candidate at $z=1.5$}
\author[M. Aravena et al.]
{\parbox{\textwidth}{M.~Aravena$^{1}$\thanks{E-mail: maravena@eso.org}, C. L. Carilli$^{2}$, 
M. Salvato$^{3}$, 
M. Tanaka$^{4}$, 
L. Lentati$^{5}$,
E. Schinnerer$^{6}$, 
F. Walter$^{6}$, 
D. Riechers$^{7}$,
V. Sm\v{o}lci\'{c}$^{8, 9, 10}$, 
P. Capak$^{7}$, 
H. Aussel$^{11}$, 
F. Bertoldi$^{9}$,
S. C. Chapman$^{12}$, 
D. Farrah$^{13,14}$,  
A. Finoguenov$^{15}$, 
E. Le Floc'h$^{11}$, 
D. Lutz$^{15}$, 
G. Magdis$^{16}$, 
S. Oliver$^{13}$,  
L. Riguccini$^{17}$
S. Berta$^{15}$,
B. Magnelli$^{15}$
and F. Pozzi$^{18, 19}$
}\vspace{0.5cm}\\
\parbox{\textwidth}{
$^{1}$European Southern Observatory, Alonso de C\'{o}rdoba 3107, Vitacura, Santiago, Chile\\
$^{2}$National Radio Astronomy Observatory, 1003 Lopezville Rd., Socorro, NM 87801, USA \\
$^{3}$Max-Planck Institut f\"ur Plasma Physik, Boltzmannstr. 2, D-85748, Garching b. M\"unchen, Germany\\
$^{4}$Institute for the Physics and Mathematics of the Universe, U. of Tokio, 5-1-5 Kashiwanoha Kashiwashi Chiba 277-8583, Japan\\
$^{5}$Cavendish Laboratory, University of Cambridge, J. J. Thompson Ave., Cambridge CB3 0HE, UK\\
$^{6}$Max-Planck Institut f\"ur Astronomie, K\"onigstuhl 17, D-69117 Heidelberg, Germany\\
$^{7}$California Institute of Technology, MC 249-17, 1200 East California Boulevard, Pasadena, CA 91125, USA\\
$^{8}$European Southern Observatory, Karl-Schwarzschild-Strasse 2, D-85748 Garching b. M\"unchen, Germany\\
$^{9}$Argelander Institute for Astronomy, Auf dem H\"ugel 71, D-53121 Bonn, Germany\\
$^{10}$University of Zagreb, Physics Department, Bijeni\v{c}ka cesta 32, 10002 Zagreb, Croatia\\
$^{11}$CNRS/Service d'Astrophysique, Bat 709 CEA-Saclay, Orme de Merisiers, F91 191 Gif-sur-Yvette, France\\
$^{12}$Institute of Astronomy, University of Cambridge, Madingley Road, Cambridge CB3 0HA, U.K.\\
$^{13}$Department of Physics and Astronomy, University of Sussex, Falmer, Brighton BN1 9QH, U.K.\\
$^{14}$Department of Physics, Virginia Tech, Blacksburg, VA 24061, USA\\
$^{15}$Max-Planck-Institut f\"{u}r Extraterrestrische Physik, Giessenbachstra$\beta$e D-85748, Garching b. M\"{u}nchen, Germany\\
$^{16}$University of Oxford, Department of Physics, Denys Wilkinson Building, Keble Road OX1 3RH Oxford, U.K.\\
$^{17}$Research Center for Space and Cosmic Evolution, Ehime University, Bunkyo-cho, Matsuyama 790-8577, Japan\\
$^{18}$Dipartimento di Astronomia, Universit\`{a} degli Studi di Bologna, Via Ranzani 1, I-40127 Bologna, Italy\\
$^{19}$INAF - Osservatorio Astronomico di Bologna, Via Ranzani 1, I-40127 Bologna, Italy
}}
\begin{document}


\maketitle
\label{firstpage}
\begin{abstract}
We report results from a deep Jansky Very Large Array (JVLA) search for $^{12}$CO $J=1-0$ line emission from galaxies in a candidate galaxy cluster at $z\sim1.55$ in the COSMOS field. We target four galaxies with optical spectroscopic redshifts in the range $z=1.47-1.59$, consistent with the likely redshift for the candidate galaxy cluster.  Two of these four galaxies, ID 51613 and ID 51813, are nominally detected in CO $1-0$ line emission at the $3-4\sigma$ level. We find CO luminosities of $(2.42\pm0.58)\times10^{10}$ K km s$^{-1}$ pc$^2$ and $(1.26\pm0.38)\times10^{10}$ K km s$^{-1}$ pc$^2$, respectively. Taking advantage from the clustering and expanded 2-GHz bandwidth of the JVLA, we perform a search for emission lines in the proximity of optical sources within the field of view of our observations ($60''$). We limit our search to galaxies with $K_\mathrm{S}<23.5$ (AB) and $z_\mathrm{phot}=1.2-1.8$. We find 2 bright optical galaxies, ID 51207 and ID 51380, to be associated with significant emission line peaks ($>4\sigma$) in the data cube, which we identify with the CO $1-0$ line emission. To test the reliability of the line peaks found, we performed a parallel search for line peaks using a Bayesian inference method. Both CO line emitting candidates are identified with probabilities of 13\% and 72\% that there are line emitting sources in each case, respectively. Monte Carlo simulations show that such associations are statistically significant, with probabilities of chance association of 3.5\% and 10.7\% for ID 51207 and ID 51380, respectively. Modeling of their optical/IR spectral energy distributions (SED) indicates that the CO detected galaxies and candidates have stellar masses and star formation rates (SFRs) in the range $(0.3-1.1)\times10^{11}\ M_\odot$ and $60-160\ M_\odot$ yr$^{-1}$, with star formation efficiencies (SFEs) comparable to that found in other star-forming galaxies at similar redshifts. By comparing the space density of CO emitters derived from our observations with the space density derived from previous CO detections at $z\sim1.5$, and with semi-analytic predictions for the CO luminosity function, we suggest that the latter tend to underestimate the number of CO galaxies detected at high-redshift. Finally, we argue about the benefits of future searches for molecular gas line emission in clustered fields with upcoming submillimeter/radio facilities.
\end{abstract}


\section{Introduction}

\begin{figure*}
\centering
\includegraphics[scale=0.6]{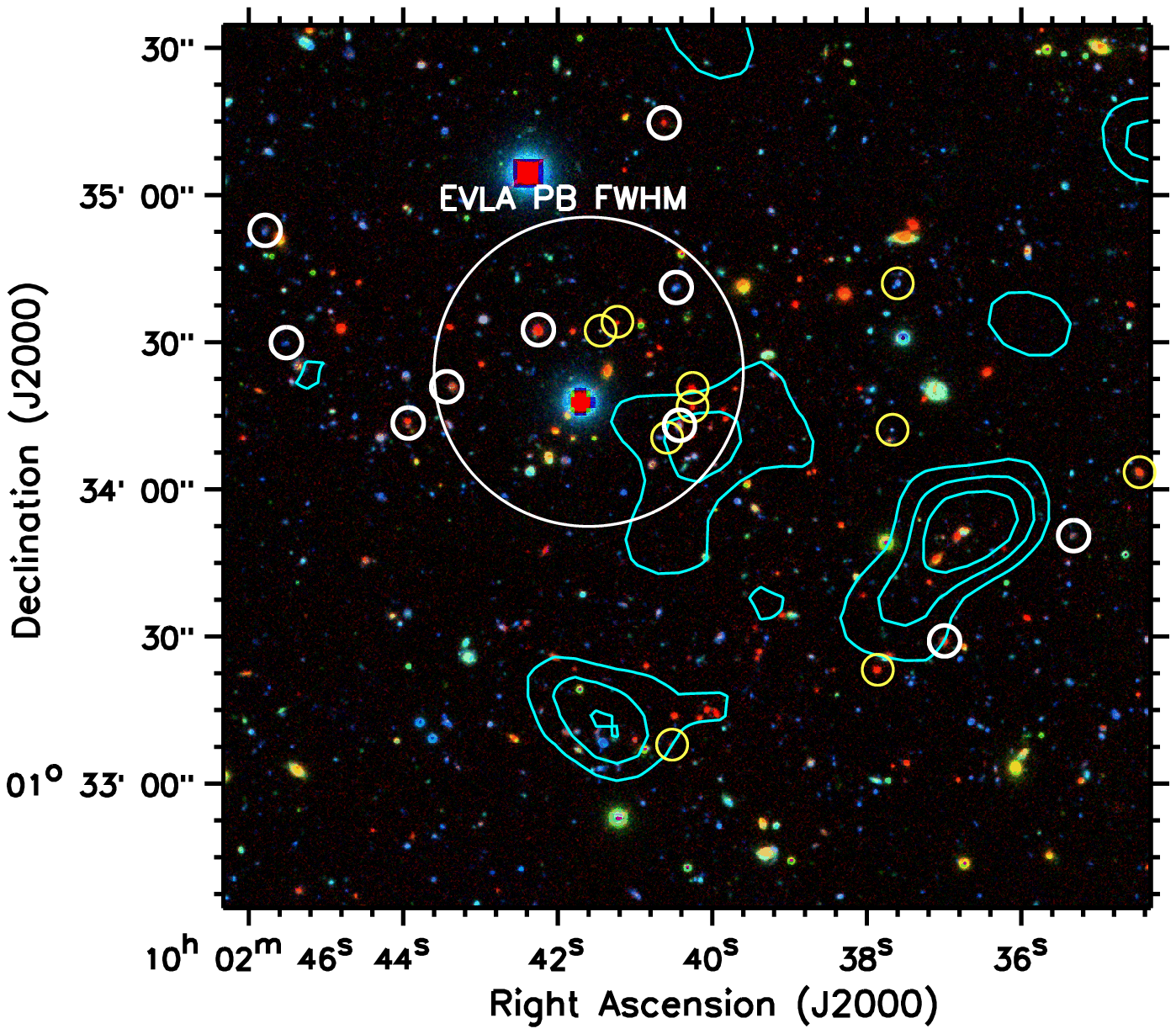}
\includegraphics[scale=0.6]{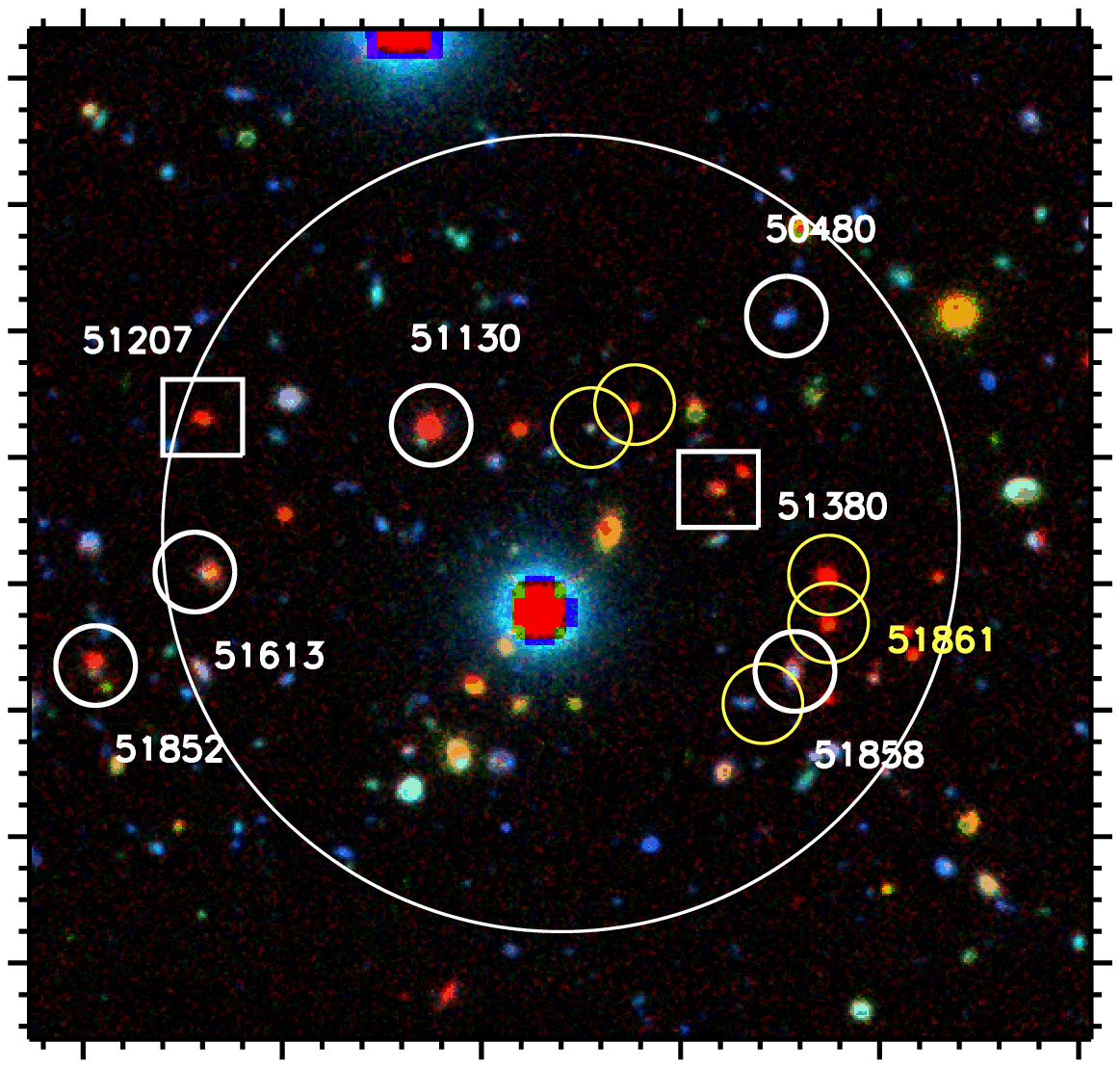}
\caption{{\it (Left:)} $Bi'K$ color composite of a $3'\times3'$ region around the candidate cluster center. Cyan contours show the XMM {\it Newton} X-ray emission at $2, 3$ and $4\sigma$ significance. White circles show spectroscopically confirmed galaxies with redshifts $z_\mathrm{spec}=1.470-1.595$. Yellow circles represent spectroscopically confirmed galaxies with $z_\mathrm{spec}=1.2-1.8$. A large $30''$ radius circle indicates the location and field of view of our deep JVLA observations of the molecular gas. {\it (Right:)} Close-up to the central $80''\times80''$ around the JVLA pointing position. Labels indicate the sources ID tags.\label{fig:field}}
\end{figure*}

One of the major goals of modern observational cosmology is to understand how the gas in the diffuse interstellar medium (ISM) of galaxies converts into stars and how both phases evolve with cosmic time. A major advance has been the determination of the evolution of the SFR density of the Universe. Deep optical and radio surveys indicated that the SFR density steadily increases from $z=5$, with a peak at $z\sim3-1$, and steeply declines from $z=1$ to the present \citep[e.g., ][]{Lilly1996, Madau1996, Cowie1999, Steidel1999, Giavalisco2004, Ouchi2004, Bouwens2007, Smolcic2009}. Although the contribution from luminous, merger-driven starburst galaxies at high-redshift to the SFR density appears to be significant, the population that drives such evolution appears to be formed by more quiescent massive star-forming galaxies \citep{Daddi2007, Rodighiero2011}. The close relationship between the SFR and the amount of molecular gas in galaxies, from which stars form, suggest that the evolution of the SFR density is the result of the evolution of the molecular gas density in galaxies across cosmic times. 

Measuring the evolution of this molecular gas density is thus critical to understand the formation of molecular gas and stars in galaxies. For this, it is necessary to perform blind surveys of the sky that allow us to measure the molecular gas content in a CO flux limited sample as a function of redshift and luminosity \citep[e.g. observations in the local Universe and simulations by ][]{Keres2003, Obreschkow2009a, Obreschkow2009b}. The technical limitations (e.g. bandwidth, sensitivity) of current submillimeter and radio facilities, along with the intrinsic faintness of the CO emission lines used to measure the amount of molecular gas in galaxies have, however, precluded such studies. The advent of sensitive interferometers such as the Jansky Very Large Array \citep[JVLA; ][]{Perley2011} and the Atacama Large Millimeter Array \citep[ALMA; ][]{Wootten2009} will make these studies possible. 

A huge step toward this goal has been the recent detection of significant CO line emission in massive star-forming galaxies at high redshift \citep{Daddi2008, Daddi2010a, Tacconi2010, Geach2011}. The large molecular gas masses derived are comparable to that observed in submillimeter galaxies (SMGs), in the range $\sim10^{10-11}\ M_\odot$, however these galaxies are forming stars at $\sim5-10$ times lower rates, with typical SFRs in the range $\sim50-300$ M$_\odot$ yr$^{-1}$. These findings imply that they have low SFEs and thus consume the gas at longer timescales, in agreement with their long duty cycle times of $>0.5$ Gyr \citep{Daddi2007}. From the analysis of the CO molecular gas and H$\alpha$ ionized gas kinematics, it is expected that between 1/3 to 2/3 of the star forming galaxies at $z\sim2$ are consistent with rotating disks \citep{ForsterSchreiber2009, Tacconi2010}, and suggest that the dominant mechanism to cool gas into molecular clouds is driven by cold gas accretion rather than galaxy interactions \citep{Keres2005, Dekel2009}. Follow-up multi-transition CO observations recently discovered that a cold gas component, with gas excitation conditions similar to that seen in local spiral galaxies, is present in these objects \citep{Dannerbauer2009, Aravena2010}. Remarkably, observations at $z=0, z=0.5$ and $z=1-2$ suggest that there has been a strong evolution in the molecular gas fractions in star forming galaxies over the past $10$ Gyr  \citep{Tacconi2010, Lagos2011, Geach2011}.

In this paper, we present results of a deep search for CO molecular gas from galaxies that are part of a candidate galaxy cluster at $z\sim1.5$ with the JVLA. The scope of this paper is to perform the first attempt to conduct an efficient blind search for CO emission from galaxies located in a part of the sky that likely contains a higher than average number of CO emitters. The  clustering of galaxies in angular scales and along the line of sight makes it possible to effectively increase the number of galaxies observed within a single pointing and frequency tuning with the JVLA. We use the extensive multi-wavelength dataset available for the COSMOS field, including the accurate photometric redshifts, to inform the location of optical galaxies and to guide the search for CO emission. In section 2, we present the properties of the (overdense) target field and the details of the JVLA observations. In section 3, we present the main results from this paper. We report the detection of CO $1-0$ emission in two spectroscopically confirmed galaxies and analyze the 45.5 GHz continuum emission and properties of a spectroscopically confirmed radio loud galaxy. In sections 3.3 and 3.4, we present the search for line emission peaks from optical galaxies in the field. Significant line peaks associated with optical sources that have compatible photometric redshifts are identified as CO line candidates. The line search is performed using two different algorithms. From sections 3.5 to 3.7, we analyze the likelihood that such associations are real based on both a statistical approach and on the detection of the CO candidates in the IR wavelength regime. In section 4, we discuss the derived SFEs for the two CO detected spectroscopically confirmed galaxies and for the two CO candidate emitters, as well as the implications of these results for future experiments to measure the CO source counts at high redshift. We adopt a concordance $\Lambda$CDM cosmology throughout, with $H_0=71$ km s$^{-1}$ Mpc$^{-1}$, $\Omega_M=0.27$, and $\Omega_\Lambda = 0.73$ \citep{Spergel2007}.

\begin{figure}
\centering
\includegraphics[scale=0.5]{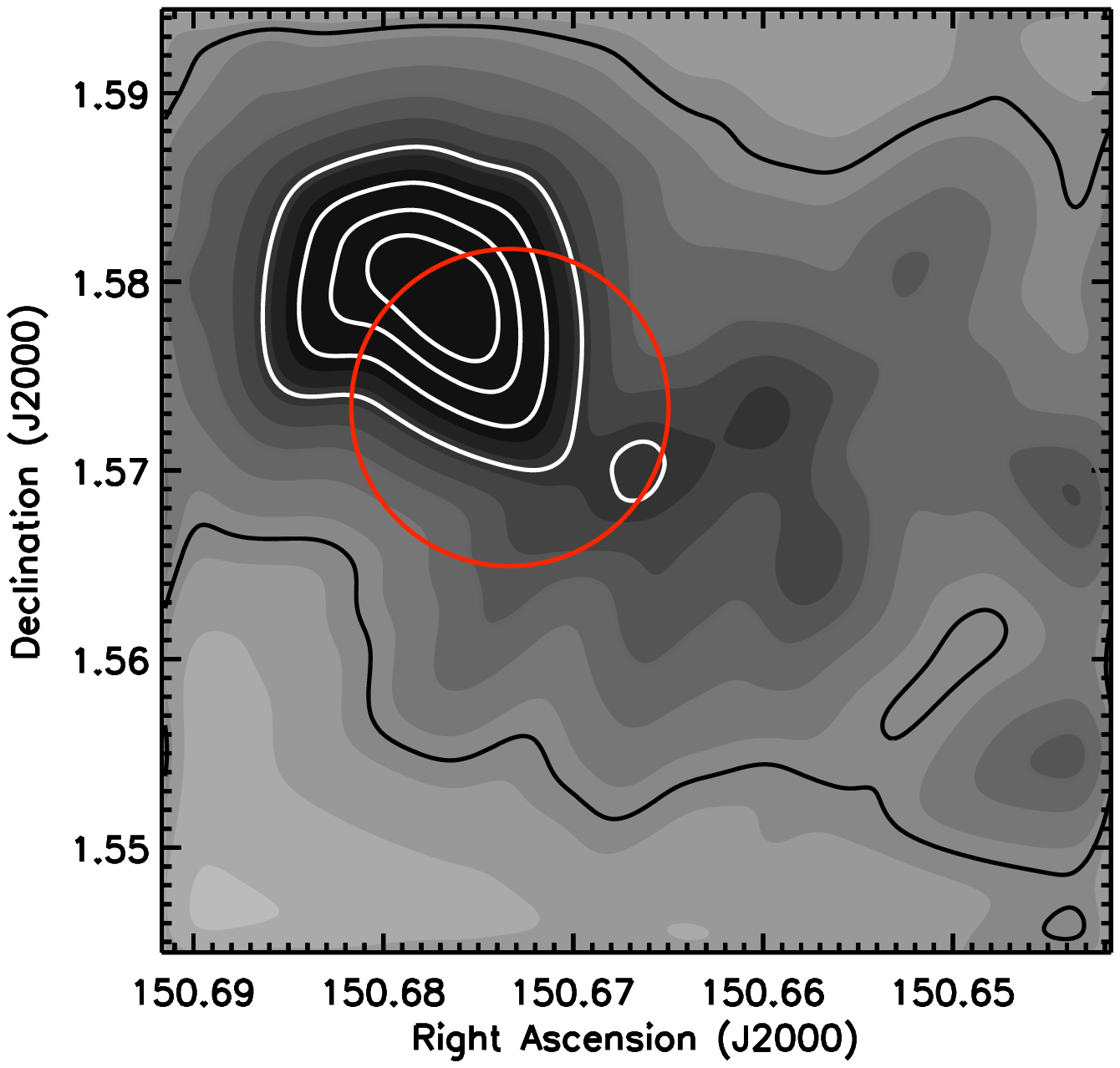}
\caption{Projected number density of galaxies with $z_\mathrm{phot}=1.5-1.6$ and SFR $>5$ M$_\odot$ yr$^{-1}$ in a $3'\times3'$ region around the cluster center. The grayscale and contours represent the density of galaxies in this field, $\delta_C$, given in terms of the average density $\delta_0$ of similarly selected galaxies in the COSMOS field, $\delta = \delta_C/\delta_0$. Contours range from 2 to 7 in steps of +1. The red circle represents the location of the JVLA pointing and primary beam (PB) FWHM. \label{fig:densitymap}}
\end{figure}

\begin{figure}
\centering
\includegraphics[scale=0.55]{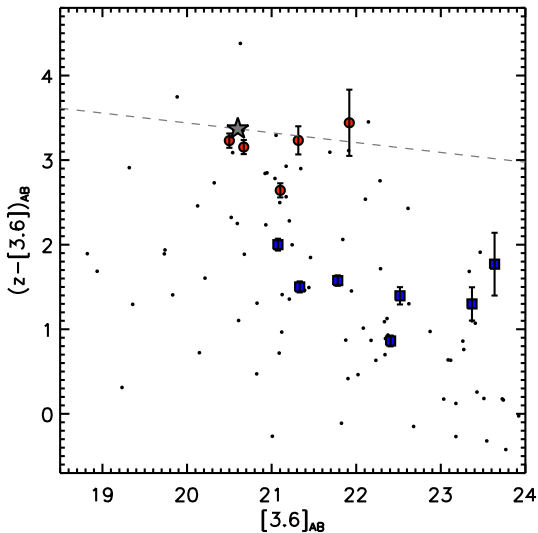}
\caption{Color-magnitude diagram ($z-[3.6]$ vs. $[3.6]$) of galaxies located within $r_{200}$ ($0.9'$) from the center of the cluster candidate. The gray dashed line shows a model red sequence for a cluster formation redshift of $z_f=3$ from \citet{Lidman2008}. The gray star symbol represents the the characteristic magnitude $m^\star$ for passively evolving galaxies. Blue filled squares correspond to galaxies in the redshift range $z=1.5-1.7$. The red filled circles represent galaxies located in the same redshift range but identified to belong to the forming red-sequence. Small black circles show background/foreground galaxies in the field. \label{fig:cm_diagram}}
\end{figure}

\begin{figure}
\includegraphics[scale=0.55]{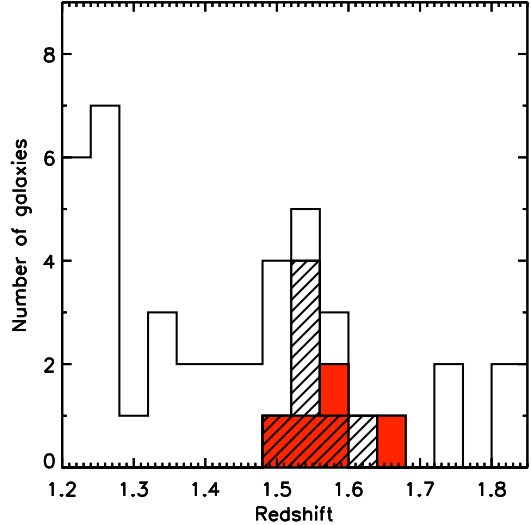}
\caption{The distribution of photometric redshifts of galaxies located within $r_{200}$ from the center of the cluster candidate. The open histogram shows all the galaxies in the field. The red histogram shows the redshift of the red-sequence identified galaxies, and the dashed histogram indicate the redshifts of galaxies shown in blue in Fig. \ref{fig:cm_diagram}. \label{fig:zdistr_cm}}
\end{figure}

\begin{table*}
\centering
\caption{Observed CO properties for galaxies with spectroscopic redshift\label{tab:1}}
\begin{tabular}{cccccccc}
\hline
ID$^a$ & $\alpha_{J2000}$ $^b$& $\delta_{J2000}$ $^b$& $z_\mathrm{opt}$ $^c$& $z_\mathrm{CO}$ $^d$& $S_\mathrm{CO}dv$ $^e$& $L'_\mathrm{CO}$ $^f$& $M(\mathrm{gas})$ $^g$ \\
 &  &   &  &   &  (Jy km s$^{-1}$) & ($10^{10}$ K km s$^{-1}$ pc$^2$) & ($10^{10}\ M_\odot$)  \\
\hline\hline
50480 & 10\ 02\ 40.47 & +01\ 34\ 41.2  & 1.523 &  -        & $<0.053$ & $<0.7$& $\ldots$ \\
51130$^{\dagger}$ & 10\ 02\ 42.25 & +01\ 34\ 32.5  & 1.519 &  -        & $<0.044$ & $<0.6$& $\ldots$ \\
51613 & 10\ 02\ 43.36 & +01\ 34\ 20.9 & 1.516 & 1.517  & $0.20\pm0.05$ & $2.42\pm0.58$ &  $8.7\pm2.1$ \\
51858 & 10\ 02\ 40.43 & +01\ 34\  13.1 & 1.560 & 1.556 & $0.10\pm0.03$ & $1.26\pm0.38$ & $4.5\pm1.4$ \\
\hline
\end{tabular}
\begin{flushleft}
\begin{footnotesize}
\noindent $^{\dagger}$ Radio galaxy; $^a$ COSMOS ID; $^b$ Position of the optical source; $^c$ Optical redshift; $^d$ CO redshift; $^e$ Spatially and velocity integrated line flux; $^f$ CO luminosity; $^g$ Gas mass obtained using $\alpha_\mathrm{CO}=3.6\ M_\odot$ (K km s$^{-1}$ pc$^2$)$^{-1}$
\end{footnotesize}
\end{flushleft}
\end{table*} 

\section{Observations}

\subsection{The COSMOS field}

COSMOS is the largest deep survey carried out with the {\it Hubble} Space Telescope (HST) covering $\sim2$ deg$^2$ in the sky \citep{Scoville2007}. Extensive imaging of the COSMOS field has been performed from the X-rays to the radio wavelengths. This includes complete optical/near-IR coverage in broad and intermediate bands with several ground-based observatories including the Subaru telescope, the Canada France Hawaii Telescope (CFHT), the United Kingdom Infrared Telescope (UKIRT), and the Kitt-Peak National Observatory (KPNO);  IR imaging with the Infrared Array Camera (IRAC) and the Multiband Imaging Photometer for Spitzer (MIPS) on board of the {\it Spitzer} Space Telescope; and radio imaging with the VLA at 20 cm. Deep imaging at far-IR wavelengths with the {\it Herschel} Space Observatory has been obtained as part of the Photodetector Array Camera and Spectrometer (PACS) Evolutionary Probe (PEP) and the {\it Herschel} Multi-tiered Extragalactic Survey (HerMES) programs \citep{Lutz2011, Oliver2010}. This has been complemented by an intensive spectroscopic campaign with ground-based observatories as part of the zCOSMOS survey \citep{Lilly2007, Lilly2009}. From the rich multi-wavelength imaging, accurate photometric redshifts were computed using 31 UV/optical/IR bands \citep{Salvato2009, Ilbert2009, Salvato2011}. These reach uncertainties of $\delta z\sim0.2-0.3$ at $z\sim1.5$ (see below). For details in the optical/IR imaging and catalogs of the COSMOS field, see \citet{Capak2007} and \citet{Ilbert2009}. The 1.4 GHz radio observations in this region of the COSMOS field reach an rms level of $30\ \mu$Jy beam$^{-1}$, and hence only sources with fluxes $>100\ \mu$Jy beam$^{-1}$ can be reliably detected. A complete description of the radio imaging is given in \citet{Schinnerer2007}, \citet{Bondi2008} and \citet{Schinnerer2010}. The targeted region lies right outside the region of the COSMOS field defined by the HST imaging coverage, and therefore it is not covered by Advanced Camera for Surveys (ACS) F814W band imaging. Hence studies of the optical morphologies of galaxies is not possible. 

\subsection{Our target: Discovery of a candidate galaxy cluster at $z\sim1.5$ with the red-sequence technique}

In this section, we present the identification of a substantial candidate galaxy cluster at $z\sim1.55$ based on four pieces of evidence: (i) the detection of a significant ($7\times$) overdensity of galaxies with photometric redshifts $z=1.5-1.6$;  (ii) the tentative detection of X-ray emission; (iii) the formation of a red sequence of a few passive galaxies in the field; and (iv) the existence of a radio loud galaxy with a confirmed spectroscopic redshift consistent with the redshift of the overdensity. We remark that regardless of whether this constitutes a virialized structure (e.g. a galaxy cluster) or not, the overdensity of galaxies within a confined redshift range allowed us to observe the CO emission from 4 spectroscopically confirmed massive galaxies in that redshift range (Fig. \ref{fig:field}), and to search for blind CO line detections from optical galaxies in the field.

Accurate photometric redshifts computed with 31 bands indicate that 26 galaxies with $K < 23$ mag (AB) and within $\sim3'$ from the central X-ray position (see below), have $z_\mathrm{phot}=1.5-1.6$. Based on their best-fit spectral energy distributions (SED), we find that 16 of the 26 galaxies are consistent with significant star-forming activity and SFR $>5\ M_\odot$ yr$^{-1}$. For galaxies in this redshift range, this represents an overdensity of $\sim7\times$ compared to field star-forming galaxies at this level. This is possibly the most compelling evidence of the existence of a cluster in this region. To illustrate this, we constructed a number density map of galaxies with photometric redshifts in the range $1.5-1.6$ and SFR $>5\ M_\odot$ yr$^{-1}$ in this region, using a procedure similar to the one introduced by \citet{Dressler1980}. For this, we created a grid of $300\times300$ positions spaced by $2''$ centered at the inferred cluster X-ray position, and computed the density from the distance to the 10th-nearest neighbour galaxy, $d_{10}$, to each grid point. The number density is thereby computed as $n_{10}=10/(\pi d_{10}^{2})$. Similar approaches have been applied to rich galaxy clusters \citep[e.g.,][]{Guzzo2007}. Figure \ref{fig:densitymap} shows the density map obtained, convolved with a Gaussian kernel that has a full-width at half maximum (FWHM) of $9''$. The number density, $\delta$, is given in terms of the average density of field galaxies with $z_\mathrm{phot}=1.5-1.6$ and SFR $>5\ M_\odot$ yr$^{-1}$ in the COSMOS field, $\delta_0 = 0.7$ arcmin$^{-2}$. The source density in this region is $\sim7\times$ larger than this average. In terms of the standard deviation in the distribution of source density in the field, $\sigma_0=0.24$ arcmin$^{-2}$, this corresponds to a fluctuation of $\sim18\sigma$ with respect to the average source density in the field. The average and standard deviation values were obtained by creating a density map of galaxies with $z_\mathrm{phot}=1.5-1.6$ and SFR $>5\ M_\odot$ yr$^{-1}$ in a nearby $\sim20'\times20'$ region that did not include the overdensity, using the same procedure as above. From the distribution of density values in this map, the average and standard deviation were computed.

The structure is tentatively detected in the XMM {\it Newton} X-ray images at the $3\sigma$ level (Fig. \ref{fig:field}), and a bright, extended radio source close to the center of the cluster is seen in the VLA 1.4 GHz images (see section 3.2). This radio source was spectroscopically confirmed to be at $z=1.5185\pm0.002$, based on the detection of the MgII, [NeV], [OII] and [NeIII] emission lines. The ubiquity of radio sources in high-redshift dense large-scale structures \citep{Miley2008}, lends support to the existence of this galaxy cluster candidate. Details of this source are given in section 3.2. 

The cluster red-sequence technique represents an effective method to select galaxy clusters. It is based on the fact that galaxy clusters typically contain a uniform population of early-type galaxies. When using two photometric band filters to bracket the location of the $4000$ \AA \ break, which is typically a strong feature in  passively evolving galaxies, early-type galaxies can be identified as the brightest and reddest objects in the field, forming a tight relation in color-magnitude diagrams. Such a ``red-sequence'' is pivoted by the brightest cluster galaxy, typically has a small scatter, and has been shown to hold out to $z\sim1.5$ \citep{Lidman2008,Bielby2010,Tanaka2010}. 

Based on the red-sequence technique modified for high-redshift \citep{Lidman2008,Wilson2009,Finoguenov2010}, we found evidence for the formation of a red sequence of passive galaxies in this field (Fig. \ref{fig:field}). We used the $(z-[3.6])$ galaxy colors, which select clusters at $z>1.2$. Figure \ref{fig:cm_diagram} shows the color-magnitude diagram obtained for galaxies located within a radius of $0.9'$ of the cluster center (0.5 Mpc at $z=1.55$). When selecting objects with photometric redshifts in the range $z_\mathrm{phot}\sim1.5-1.7$, a number of bright, red galaxies compared to the bulk of ``blue'' galaxies is visible. Figure \ref{fig:zdistr_cm} shows the distribution of photometric redshifts for galaxies in the field, highlighting the distribution formed from the red and blue galaxies from Fig. \ref{fig:cm_diagram}. From the redshift distribution for all galaxies in this region (empty histogram), it can be seen that there is a redshift peak in the distribution at $z\sim1.5-1.6$.

In addition to the available optical spectroscopic redshifts provided by zCOSMOS, a dedicated spectroscopy campaign targeting galaxies within a $3'\times3'$ region around the center of the cluster candidate was performed using the Deep Imaging and Multi-Object Spectrograph (DEIMOS) on the Keck telescope and Multi-Object Infrared Camera and Spectrograph (MOIRCS) on the Subaru telescope (Fig. \ref{fig:field}; Tanaka et al. in preparation). Due to bad weather conditions, however, it was not possible to obtain redshifts of the passive galaxies in the field. Nine sources within this region were confirmed to have spectroscopic redshifts in the range $1.47-1.59$, which corresponds approximately to the range covered by the JVLA observations (see next section; Fig. \ref{fig:field}), including the galaxy associated with bright 1.4 GHz radio emission. This redshift range represents a large range in velocity, $>10000$ km s$^{-1}$, which could point to a filamentary structure nature for the overdensity rather than a collapsed one. 

Based on the spectroscopic and photometric redshift evidence, we pointed our JVLA observations targeting four spectroscopically confirmed galaxies that were located close to the maximum galaxy density in this field. In order to observe the CO $1-0$ emission from the four spectroscopically confirmed galaxies, which includes the bright radio source, we tuned the JVLA receivers to cover the redshift range $z\sim1.47-1.59$ (see next section). Hereafter, and in the absence of spectroscopic evidence confirming this structure, we refer to this as a cluster candidate or overdensity. We stress that the existence of an overdensity in photometric redshift and angular scales, added to the large bandwidth of the JVLA, enables us to search for CO $1-0$ emission from galaxies in this field irrespective of whether this overdensity is a cluster or not.

\subsection{JVLA observations}

\begin{figure*}
\centering
\includegraphics[scale=0.42]{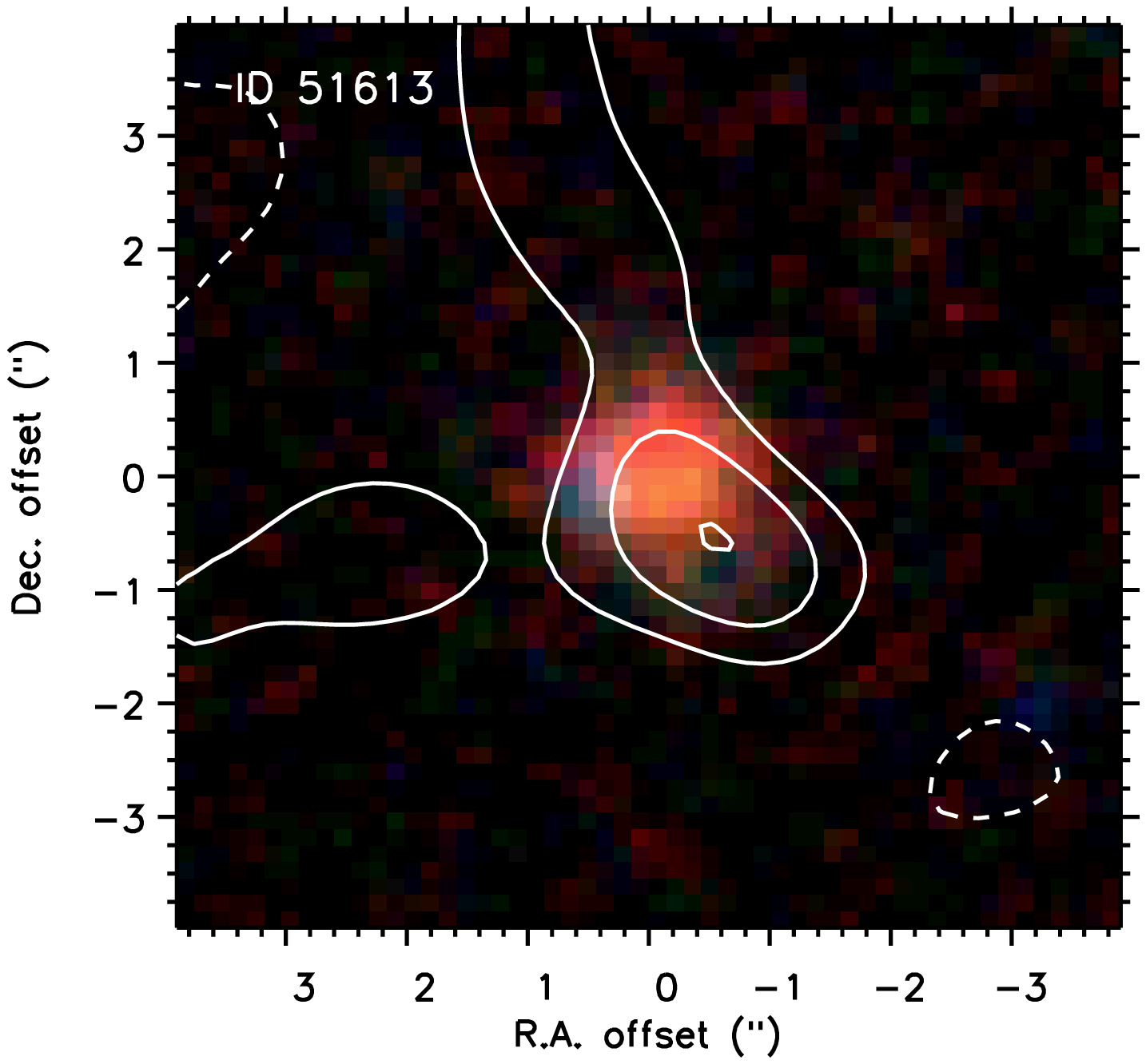}
\includegraphics[scale=0.42]{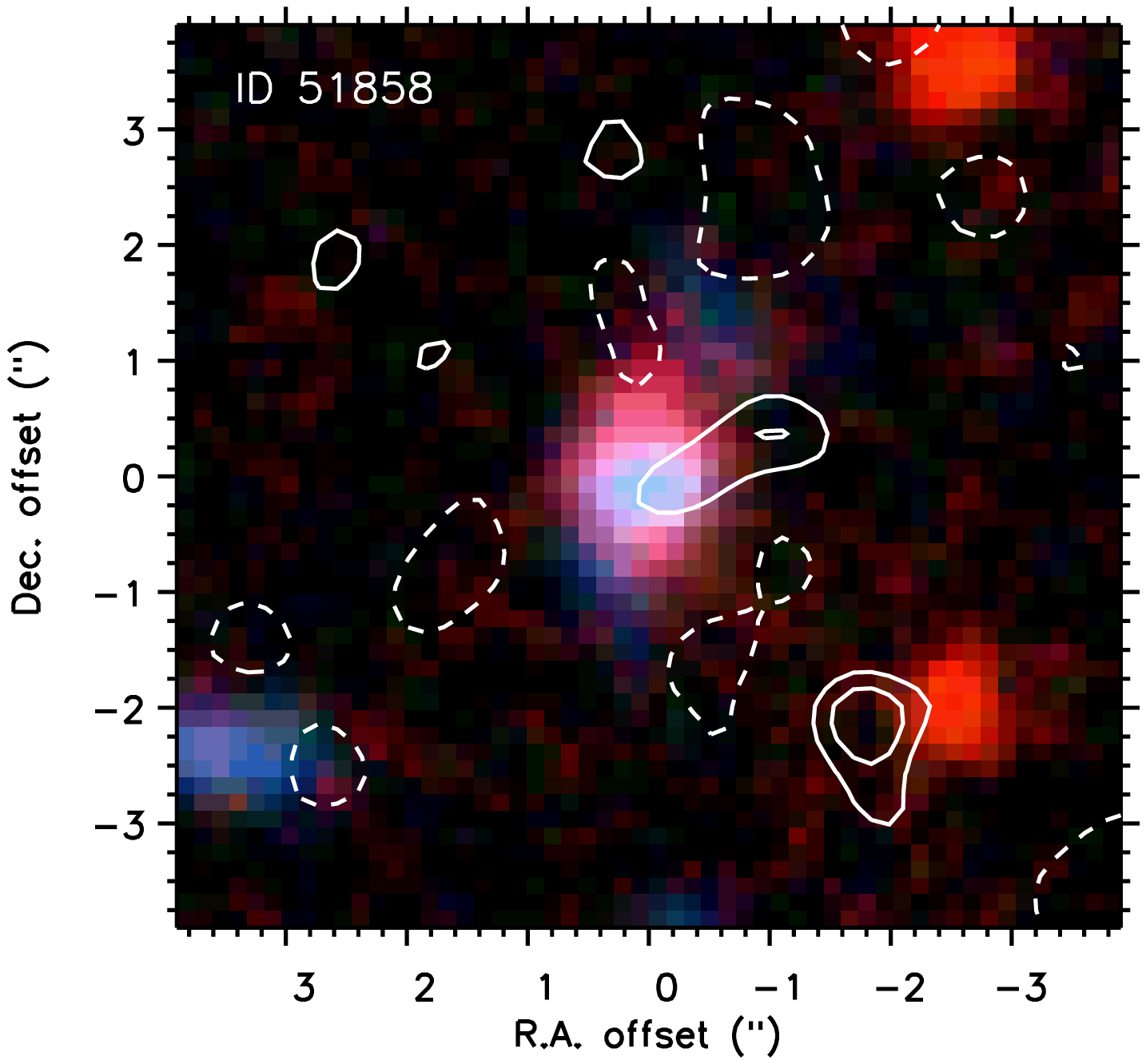}
\caption{JVLA CO $1-0$ emission line intensity map (PB corrected), averaged over 230 and 500 km s$^{-1}$ and centered at 45.810 GHz and 45.107 GHz for ID 51613 and ID 51858, respectively. The background corresponds to a $Bi^+K$ color-composite image. The CO emission is shown in contours in steps of $1\sigma$, starting at $\pm2\sigma$, with $\sigma=120$ and 40 $\mu$Jy beam$^{-1}$, for  ID 51613 and ID 51858, respectively. The corresponding beam sizes are $2.1''\times1.9''$ and $1.1''\times1.0''$, respectively. Dashed contours represent negative signal. In the case of ID 51613 the resolution has been downgraded to increase the signal to noise. \label{fig:specmaps}}
\end{figure*}

\begin{figure*}
\centering
\includegraphics[scale=0.49]{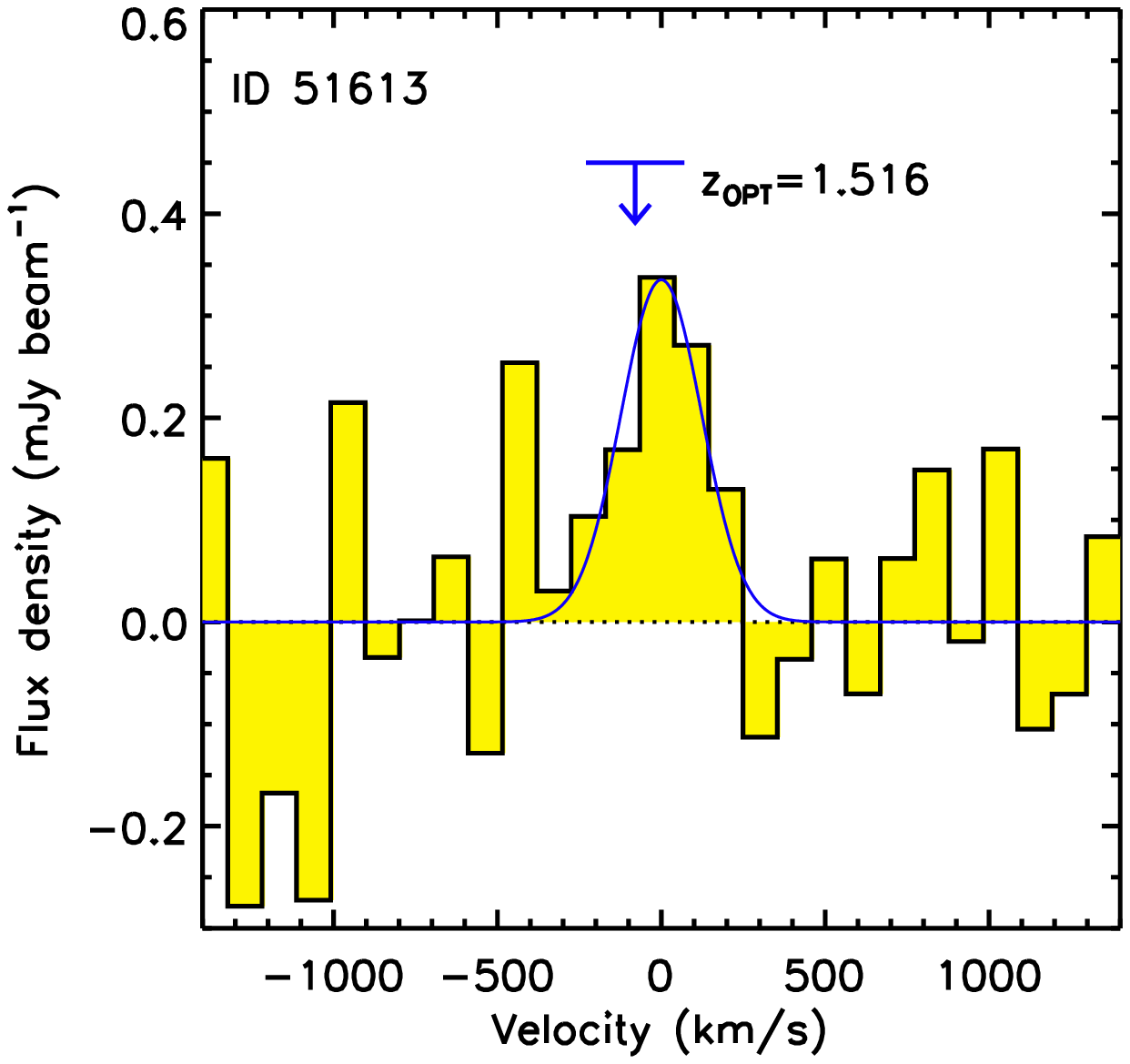}
\includegraphics[scale=0.49]{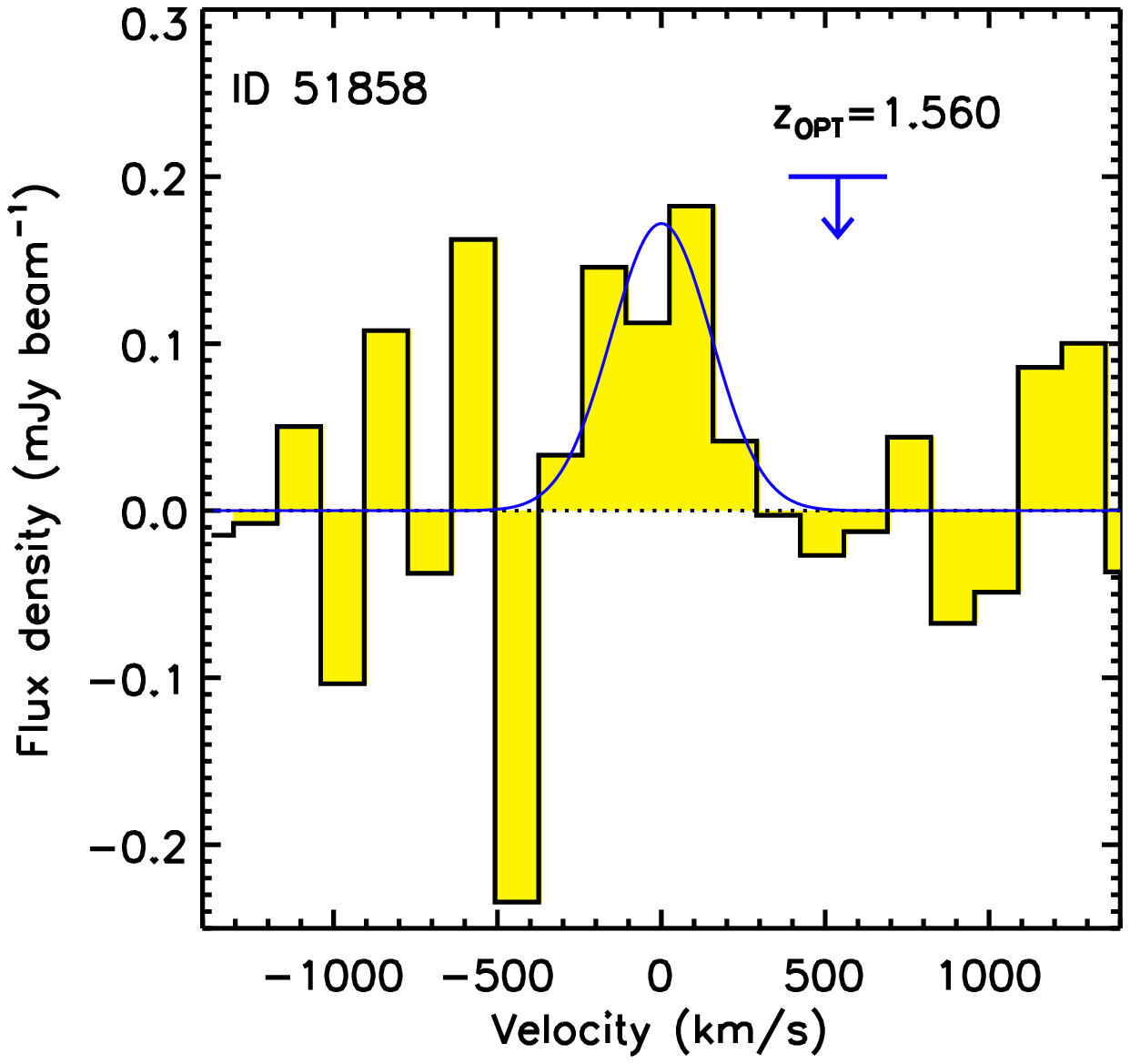}
\caption{JVLA 45.5 GHz close-up spectrum towards the optical galaxies ID 51613 and ID 51858 at 93 and 132 km s$^{-1}$ per channel resolution, respectively. The emission profiles were measured at the peak position of the integrated emission close to the optical source. The arrow represents the frequency that corresponds to the CO $1-0$ line at the optical spectroscopic redshift. The emission features identified with the CO 1-0 line emission are highlighted by a Gaussian fit (blue solid line). \label{fig:specspec}}
\end{figure*}

Observations of the $^{12}$CO $J=1-0$ emission line ($\nu_\mathrm{rest}=115.271$ GHz) in the cluster galaxies at $z\sim1.5$ were taken as part of the JVLA Residence Shared Risk Observing (RSRO) program (program ID: AA330).  The target field was observed during March 2010 to January 2011 and centered at $\alpha_{J2000}=10^h02^m41.6^s$, $\delta_{J2000}=+01^\circ34^{'}24.^{''}0$. The observations were made using the Q-band receivers (tunable range: 40-50 GHz) and taking advantage of the expanded bandwidth of the WIDAR correlator ($\approx2$ GHz).   

Based on early spectroscopic and photometric redshift estimates that indicated a cluster redshift of $z\approx1.53$, we tuned our observing frequency to 45.500 GHz. We used a wide bandwidth mode, with 2 intermediate frequency (IF) bands (or basebands: AC and BD), each formed by 8 spectral windows (SPW). Each SPWs was composed of 256 channels with a resolution of 2 MHz per channel. The SPW IDs range from 0 to 15. Combined, they cover a total bandwidth of 2048 MHz ($\sim2$ GHz). Since each SPW loses sensitivity at the edge channels and because at the time when the observations were taken the WIDAR correlator could not be configured to overlap different SPWs, we used two 2048 MHz bandwidth spectral setups shifted by 5 channels from each other. Two channels at each edge of the SPWs were thus flagged. This avoids spectral gaps, obtaining an even frequency coverage. At the edge SPW of each IF, the passband deteriorated more than in the other SPWs. Thus, 5, 3, 4 and 4 additional channels at the external IF borders of SPW ID 0, 7, 8, and 15, respectively, were flagged. This results in a coarser frequency coverage and a spectral gap of 8 channels (16 MHz or 105 km s$^{-1}$) in the middle of the band. Such problem was unforeseen prior or during the observations. Hence, the observations covered the frequency ranges $44.486-45.492$ GHz (AC) and $45.508-46.514$ GHz (BD), or approximately 1 GHz per band. This implies that we are able to detect the CO $1-0$ line emission from galaxies in the redshift range $z=1.478-1.591$ with a negligible gap of $\Delta z\approx0.0009$.

Although for detection experiments the D-array configuration is better suited, the observations were taken in C-array configuration due to scheduling constraints. This implies that sources extended by more than $1''$\ were slightly resolved with the consequent loss in sensitivity. The nearby source J1038+0512 ($\sim0.5-0.7$ Jy; $\sim9^\circ$ distance to target) was observed every $\sim10$ min for gain calibration. The brighter sources J1256-0547 (3C279; $\sim16-21$ Jy) and J1331+3030 (3C286; $\approx1.4$ Jy) were used for bandpass and flux calibration, respectively. Pointing calibration was performed on J1038+0512 every $\sim55$ min using the C-band receivers at 4.9 GHz. 

Data calibration was done using the Common Astronomy Software Application (CASA) package. Time ranges with poor visibilities as well as the edge channels where the bandpass deteriorated, as explained above, were flagged. The final integration time amounts to $\sim20$ hrs on-source. At the observing frequency, the JVLA primary beam (PB) FWHM is $\sim60''$. To obtain flux measurements, we deconvolved the visibilities using the CLEAN task with natural weighting and a further taper of $0.8''$, and applied the corresponding primary beam correction. We thus obtained a final resolution of $1.1''\times1.0''$ and an average rms of $60$ $\mu$Jy beam$^{-1}$ per 100 km s$^{-1}$ channel. This corresponds to a physical scale of $\sim8.5$ kpc at the redshift of the overdensity and thus well suited to extract their extended emission without losing many long baselines.

\section{Results}

\subsection{CO $1-0$ in two spectroscopically confirmed galaxies}

Our JVLA observations were designed to cover the CO $1-0$ line emission from four galaxies that have an accurate optical/near-IR spectroscopic redshift in the range $z=1.478-1.591$ and located spatially within the JVLA PB FWHM.

Since the frequency of the CO $1-0$ line emission could be slightly offset from the frequency implied by the optical redshift, we first searched for a CO emission peak in the spectra along the line of sight. From this, we estimated the velocity integrated line flux emission at the CO peak frequency, if any. If no emission was detected in the spectra at the source position, we measured a limit to the emission based on the optical redshift assuming a typical line FWHM of 300 km s$^{-1}$. 

Two out of the 4 galaxies with optical spectroscopic redshift, ID 51858 and ID 51613, were nominally detected in CO $1-0$ emission with significances of $\approx3\sigma$ and $\approx4\sigma$, respectively.  Figures \ref{fig:specmaps} and \ref{fig:specspec} show the velocity averaged CO emission maps and the CO line profiles, respectively, computed at the position of the CO image peak emission. Table \ref{tab:1} summarizes the positions, redshifts, physical properties and derived parameters for these sources. 

After inspection of the CO images with $1.1''\times1.0''$ resolution, the emission from ID 51613 appeared to be slightly resolved along the major axis. In order to perform flux measurements in this case, we applied a taper of $1.7''$ to the visibilities, leading to an image resolution of $2.15''\times1.90''$.

From a Gaussian fit to the spectral line profiles, we measured formal line widths of $360\pm220$ km s$^{-1}$ and $200\pm80$ km s$^{-1}$ at central frequencies of 45.1087 GHz and 45.8030 GHz for ID 51858 and ID 51613, respectively. Spatially integrated flux densities were obtained using a two-dimensional Gaussian fit to the CO maps. These results imply CO redshifts of 1.5554$\pm$0.0008 and 1.5167$\pm$0.0004 for ID 51858 and ID 51613, respectively. This leads to a difference of $\sim590$ km s$^{-1}$ with respect to the optical redshift in the former case. Such difference is large compared to that observed in some of the massive disk galaxies reported by \citet{Daddi2010a}, of $\sim200-300$ km s$^{-1}$, however it is within the uncertainties implied by the low significance of this detection and the accuracy of the optical spectroscopic redshift determination, typically $\delta z \sim0.001$, and well within the typical velocity differences between CO observations and rest-frame UV for SMGs and QSOs, with a typical range $0-500$ km s$^{-1}$ \citep{Greve2005}.

\begin{figure}
\centering
\includegraphics[scale=0.5]{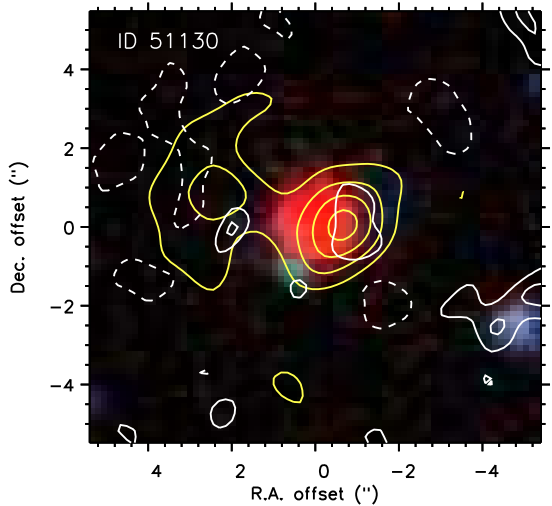}
\caption{Radio continuum emission towards the massive radio loud galaxy ID 51130. A $Bi^+K_S$ color composite image is shown in the background. The VLA 1.4 GHz emission is shown in yellow contours with levels in steps of $2\sigma$, starting at $\pm2\sigma$, with $\sigma=30\ \mu$Jy beam$^{-1}$. The 45.5 GHz continuum emission is represented by white contours in steps of $1\sigma=10\ \mu$Jy beam$^{-1}$, starting at $\pm2\sigma$. Dashed contours represent negative fluctuations.\label{fig:radiogx}}
\end{figure}

\begin{figure}
\centering
\includegraphics[scale=0.5]{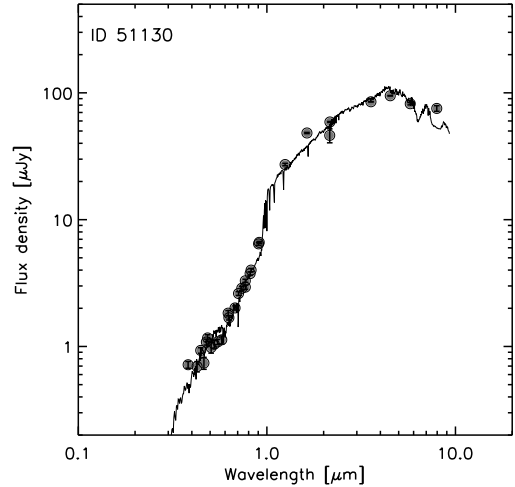}
\caption{Optical to mid-IR SED of ID 51130. Gray filled circles with error bars represent the multi-wavelength photometry and the solid line shows the best fit template at the source's spectroscopic redshift $z=1.519$.\label{fig:radiogxsed}}
\end{figure}

\begin{figure}
\centering
\includegraphics[scale=0.5]{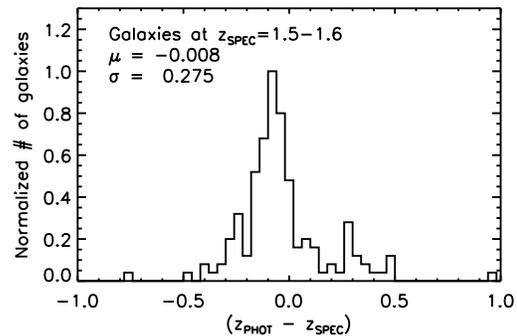}
\caption{Comparison between spectroscopic and photometric redshifts for galaxies in the COSMOS field at $z=1.5-1.6$. A dispersion $\sigma=0.275$ is measured from the mean $\mu=-0.008$.\label{fig:zcomp}}
\end{figure}

\subsection{A massive radio galaxy at $z=1.519$}

Figure \ref{fig:radiogx} shows the VLA 1.4 GHz image of a prominent double peaked radio source identified in the field (see also Fig. \ref{fig:field}). The radio source is associated with the bright optical galaxy ID 51130, which has an optical spectroscopic redshift of 1.519 (Table \ref{tab:1}). Both 1.4 GHz peaks are identified as the source's radio lobes. 

Based on the presence of radio lobes, a strong contribution from AGN activity would be expected. Since the source is not detected in the XMM {\it Newton} X-ray maps, we explored the shape of the mid-IR SED since AGN sources tend to have a steeply rising SED from 3.6 $\mu$m to 24 $\mu$m. For this, we compare the mid-IR colors of ID 51130 with three different color-color criteria from \citet{Lacy2004}, \citet{Stern2005} and \citet{Ivison2004}. ID 51130 has flux densities $S_{3.6\mu\mathrm{m}}=87\ \mu$Jy, $S_{4.5\mu\mathrm{m}}=97\ \mu$Jy, $S_{5.8\mu\mathrm{m}}=84\ \mu$Jy, $S_{8.0\mu\mathrm{m}}=78\ \mu$Jy and $S_{24\mu\mathrm{m}}=220\ \mu$Jy, respectively, being clear it has a relatively flat mid-IR SED (Fig. \ref{fig:radiogxsed}). While, the criterion proposed by Lacy et al. selects AGN based {\it roughly} on the colors log($S_{5.8\mu\mathrm{m}}/S_{3.6\mu\mathrm{m}}$)$>-0.1$ and log($S_{8.0\mu\mathrm{m}}/S_{4.5\mu\mathrm{m}}$)$>-0.2$, Stern et al. uses limits roughly defined by $0.7<(m_{5.8}-m_{8.0})<1.6$ and $(m_{3.6}-m_{4.5})>0.3$. Similarly, Ivison et al. defines a region occupied by AGN with $S_{8.0\mu\mathrm{m}}/S_{4.5\mu\mathrm{m}}>1$ and $S_{24\mu\mathrm{m}}/S_{8.0\mu\mathrm{m}}<5$. Based on this, we find that ID 51130 only satisfies the Lacy et al. AGN criterion. The source is located outside the AGN regions defined by Stern et al. and Ivison et al., however in both cases this could still be consistent with a Seyfert 2 galaxy. From this, we conclude that there is not a substantial contribution from an AGN to the galaxy's emission at optical/IR wavelengths.

Based on the lack of evidence pointing to an important AGN contribution, we fitted a pure galaxy model (with no AGN contribution) to the source's photometry (Fig. \ref{fig:radiogxsed}; see section 3.7). Note that fitting pure galaxy templates to objects with no substantial AGN contribution such as Seyfert 2 galaxies is consistent with results from \citet{Salvato2011}. The host galaxy's SED is best fitted by an 0.5 Gyr old galaxy template with a SFR of 25 M$_\odot$ yr$^{-1}$ that appears to have already formed a good fraction of its stars, as suggested by the large stellar mass (Table 3). This is consistent with the non-detection of CO line emission from this object down to a limit of $L'_\mathrm{CO}=0.6\times10^{10}$ K km s$^{-1}$ pc$^2$ (3$\sigma$). Since the nature of the host galaxy is unknown, we assume a CO luminosity to gas mass conversion factor $\alpha_\mathrm{CO}=3.6$ M$_\odot$ (K km s$^{-1}$ pc$^2$)$^{-1}$, which is the typical value found for disk galaxies at high-redshift \citep{Daddi2010a, Magdis2011}, and derive a limit to the gas mass of this object of $2\times10^{10}$ M$_\odot$ ($3\sigma$). We notice that this value of $\alpha_\mathrm{CO}$ traces the total amount of gas, including H$_2$, Helium and HI, with the vast majority of the hydrogen ($>90\%$) in molecular form \citep[see][]{Daddi2010a}. Using a conversion factor of $\sim1$ M$_\odot$ (K km s$^{-1}$ pc$^{2}$)$^{-1}$ would only put a stringent but less conservative limit. This limit is also consistent with the molecular gas masses typically found for radio AGN in the nearby Universe \citep[e.g. Table 4 in ][]{Smolcic2011}.

To measure the continuum emission of this object at 45.5 GHz, we averaged the whole 2 GHz data set over channels and created a continuum image, inverting the visibilities using natural weighting with an additional tapering of $0.8''$. Figure \ref{fig:radiogx} compares the optical view of this galaxy with the radio emission observed at 1.4 GHz and at 45.5 GHz. The 1.4 GHz radio maps reach an rms level of $\sim30\ \mu$Jy beam$^{-1}$ in this region of the COSMOS field. Two positive blobs, close to where the bulk of 1.4 GHz emission is located, can be distinguished in the 45.5 GHz continuum images, although at very low significance, $\sim2-3\sigma$. By fitting a double component two-dimensional Gaussian to the 45.5 GHz map, we find an integrated flux density of $145\pm50$ $\mu$Jy and $23\pm8$ $\mu$Jy for the western and eastern radio peaks, respectively. At 1.4 GHz, both peaks are detected with integrated fluxes of $400\pm30$ $\mu$Jy and $520\pm90$ $\mu$Jy, respectively \citep{Schinnerer2007, Bondi2008, Schinnerer2010}. Deeper observations at 45.5 GHz would be needed to properly measure the spectral index of the radio lobes.

\subsection{Search for CO line emission in galaxies}

\begin{table*}
\centering
\caption{Observed line properties for CO-identified galaxies with photometric redshift in the range $1.2-1.8$\label{tab:2}}
\begin{tabular}{ccccccccc}
\hline
ID$^a$ & $\alpha_{J2000}$ $^b$& $\delta_{J2000}$ $^b$& $d$ & $z_\mathrm{phot}$ $^c$& $z_\mathrm{CO}$ $^d$& $S_\mathrm{CO}dv$ $^e$& $L'_\mathrm{CO}$ $^f$& $M(\mathrm{H}_2)$ $^g$\\
& (J2000) & (J2000) &   ($''$) &  &   &  (Jy km s$^{-1}$) & ($10^{10}$ K km s$^{-1}$ pc$^2$) & ($10^{10}\ M_\odot$)  \\
\hline\hline
51207 &  10$^h$02$^m43.400^s$ & $+01^{\circ}34'33.12''$ & 0.60 & 1.529 &  1.530 & $0.085\pm0.021$  &  $1.03\pm0.25$ &  $3.7$ \\
51380 &  10$^h$02$^m40.810^s$ & $+01^{\circ}34'27.51''$ & 0.34 & 1.450 &  1.551 &  $0.090\pm0.021$ &  $1.12\pm0.26$  &  $4.0$ \\
\hline
\end{tabular}
\begin{flushleft}
\begin{footnotesize}
\noindent $^a$ COSMOS ID; $^b$ Position of the optical source; $^c$ Optical photometric redshift; $^d$ CO redshift; $^e$ Spatially and velocity integrated line flux; $^f$ CO luminosity; $^g$ Gas mass obtained using a galactic conversion factor $\alpha_\mathrm{CO}=3.6\ M_\odot$ (K km s$^{-1}$ pc$^2$)$^{-1}$.
\end{footnotesize}
\end{flushleft}
\end{table*} 

We took advantage of the unprecedented multi-wavelength coverage of the COSMOS field, as well as the accuracy of photometric redshifts computed with 31 UV/optical/IR bands \citep{Ilbert2009, Salvato2011}, to identify possible CO emission line peaks from optical galaxies within the field of view ($60''$). 

We selected galaxies with $K_S<23.5$ mag (AB) and photometric redshifts in the range $1.2-1.8$. The magnitude limit is set to guarantee that the photometric redshift determination is accurate, based on the comparison between spectroscopic and photometric redshifts for the COSMOS field; and that we trace galaxies with stellar masses $>2\times10^9$ M$_\odot$.  As mentioned above, our JVLA spectroscopic observations trace the CO $1-0$ emission line in the redshift range $\approx1.48-1.59$. In this range, the photometric redshift accuracy is $\Delta z=\left| z_\mathrm{spec}-z_\mathrm{phot}\right|\approx0.275$, with a small systematic offset $z_\mathrm{spec}\approx z_\mathrm{phot}+0.008$ (Fig. \ref{fig:zcomp}). Hence, to comprise most of the galaxies in the redshift range $1.48-1.59$, we conservatively selected galaxies with photometric redshifts in the range $1.2-1.8$. Based on the spectroscopic redshifts, we expect 5-10\% of galaxies to be photometric redshift outliers. From this, we have that a total of 26 galaxies are thus selected, of which up to 3 are expected to lie outside the specified redshift range.

For each selected optical source, we extracted a $10''\times10''$ cutout from the full JVLA data cube, and searched for significant ($>4\sigma$) emission line peaks lying within a $1.0''$ \ radius from the optical source position.  We performed the search making use of the AIPS task SERCH. For this, we used a data cube that had been de-convolved using a {\it Briggs} weighting scheme, with a robust parameter of 1.0, and an additional tapering of $0.8''$, without PB correction. The weighting scheme without PB correction ensures a similar noise behavior across the image, for all source positions. The SERCH task uses a Gaussian kernel to convolve the data cube along the frequency axis with an expected input line width, and reports all channels and pixels having a signal to noise ratio over the specified limit. We experimented with various Gaussian kernel line widths, from $\sim200$ to 500 km s$^{-1}$ and found that an optimal width is using $\sim350$ km s$^{-1}$. Using larger line widths does not increase the number of coincidences between optical sources and line peaks, however smaller widths ($\sim200$ km s$^{-1}$) tend to produce a significantly larger number of line peaks ($30\%$ more) with no optical associations, which are likely spurious.

A similar approach to search for CO candidates at high redshift is to blindly search for peaks in the full field-of-view of the JVLA observations, and then cross-correlate the possible detections with the optical catalog, using the same constraints on magnitudes and photometric redshifts and searching for nearby associations. Since such approach is still limited by the existence of optical sources, it essentially leads to the same results. We focused only on line peaks associated to relatively bright optical sources since this approach will produce more reliable results. Note that the frequency coverage of our observations also enables us to detect CO $2-1$ emission from galaxies in the redshift range $z=3.956-4.182$, however such sources are expected to be very faint in the optical, being more difficult to find a clear optical counterpart identification. Search for line peaks associated to $z\sim4$ sources is out of the scope of this paper.

\begin{figure*}
\includegraphics[scale=0.42]{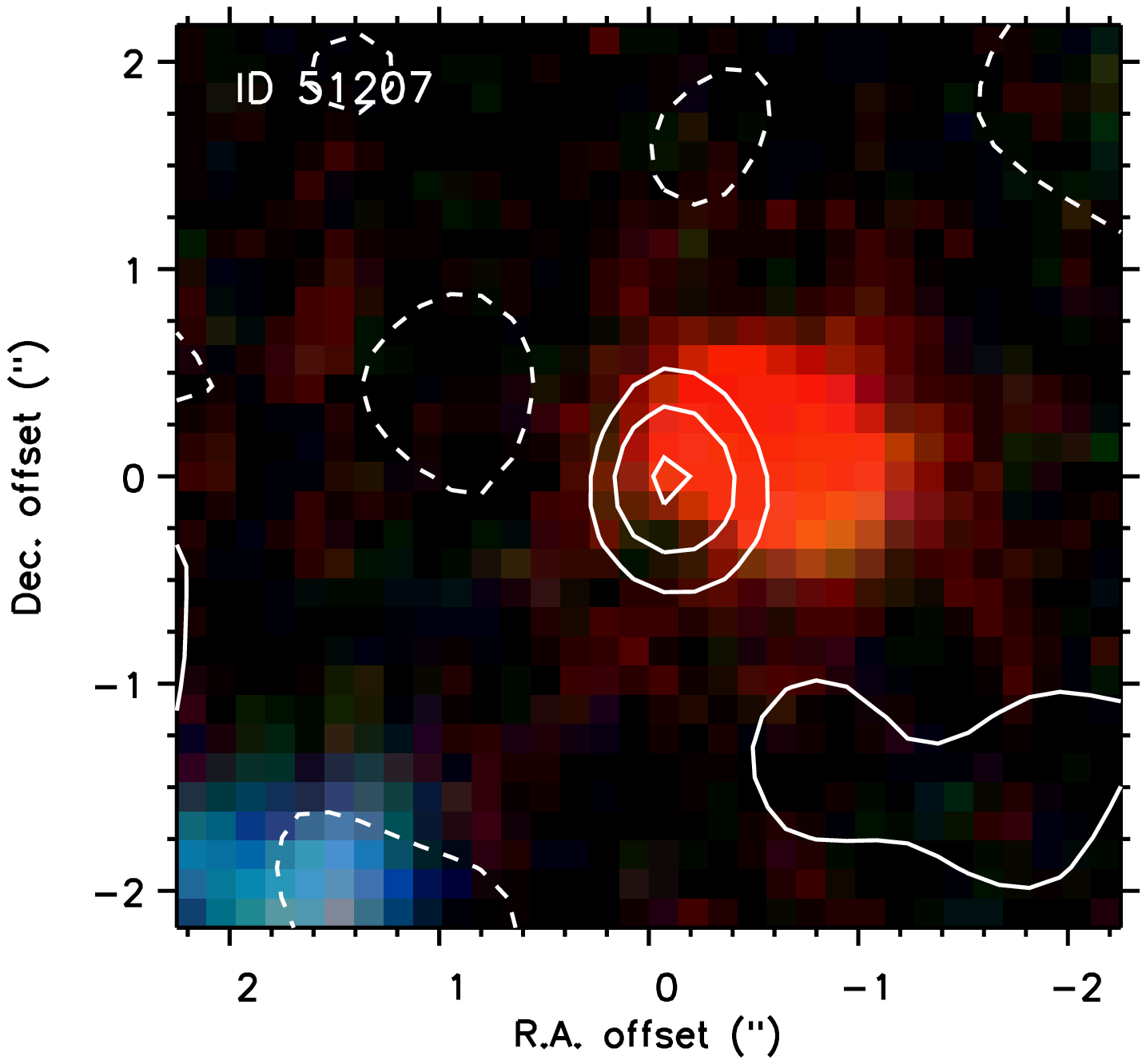}
\includegraphics[scale=0.42]{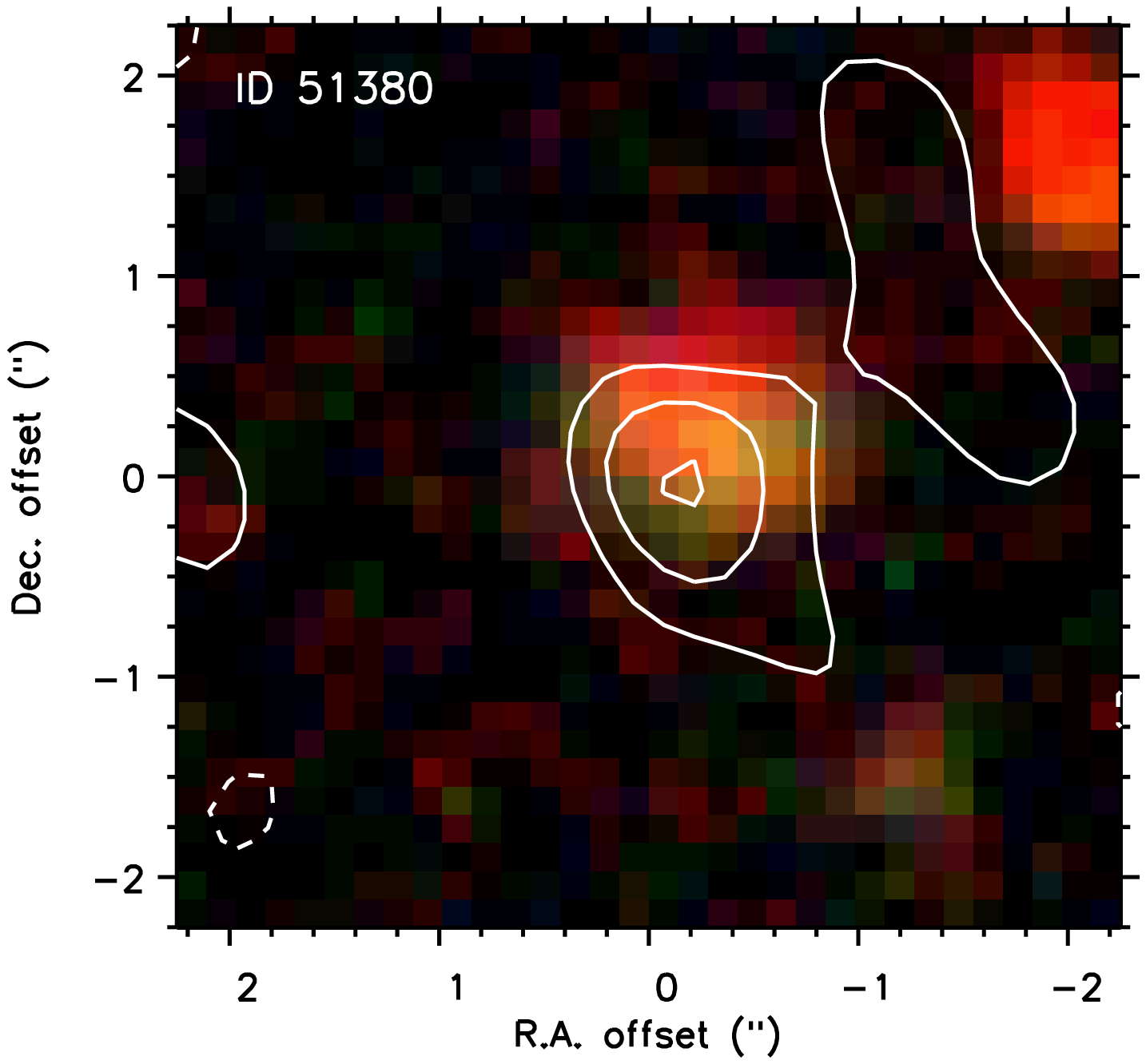}
\caption{JVLA CO $1-0$ emission line intensity map (PB corrected), averaged over 480 and 260 for ID 51207 and ID 51380, respectively. The background image corresponds to a $Bi^+K$ color-composite. The candidate CO emission is shown in contours in steps of $1\sigma$, starting at $\pm2\sigma$, with $\sigma=53$ and $57$ $\mu$Jy beam$^{-1}$, respectively.\label{fig:specmaps2}}
\end{figure*}

\begin{figure*}
\centering
\includegraphics[scale=0.49]{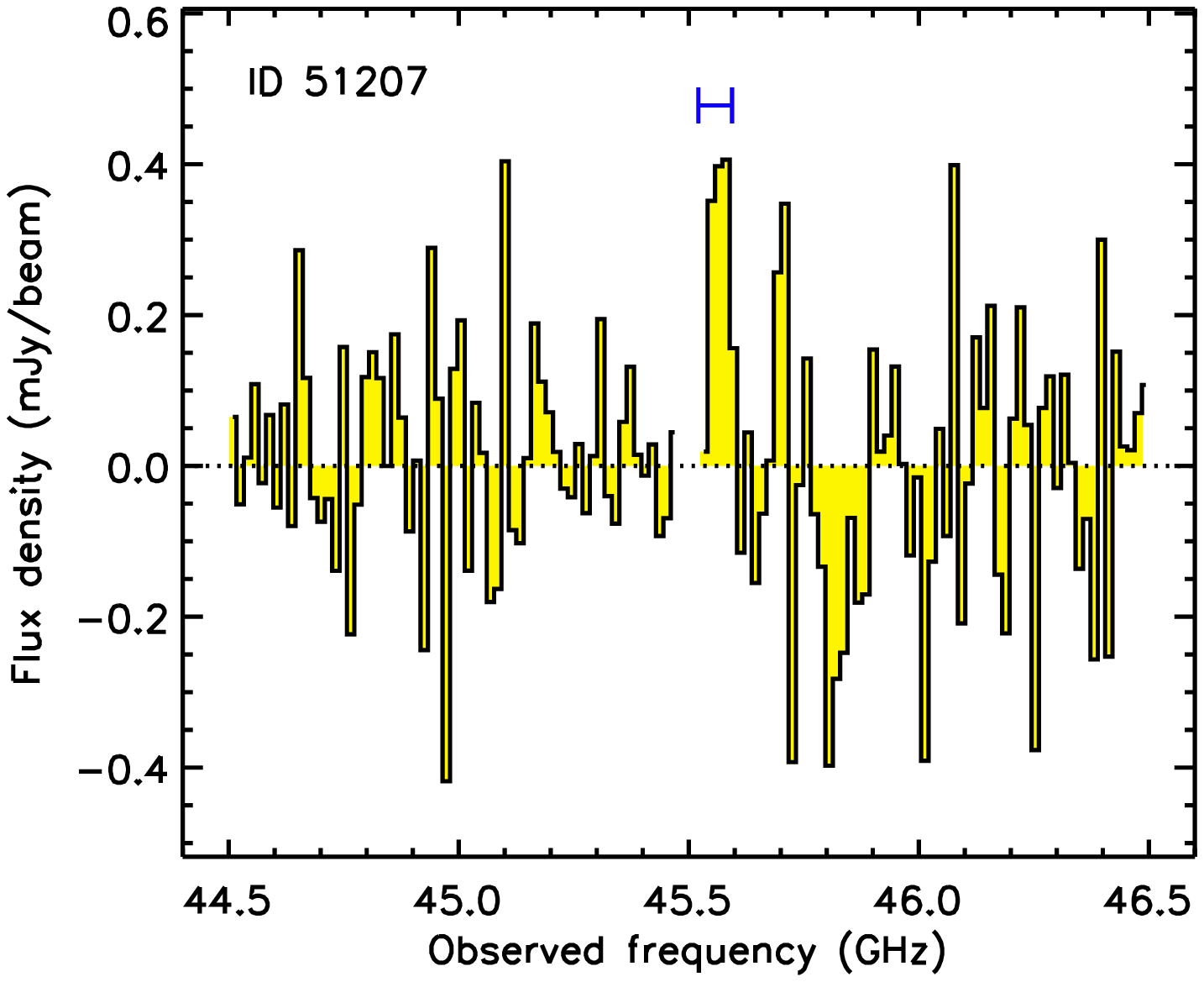}
\includegraphics[scale=0.49]{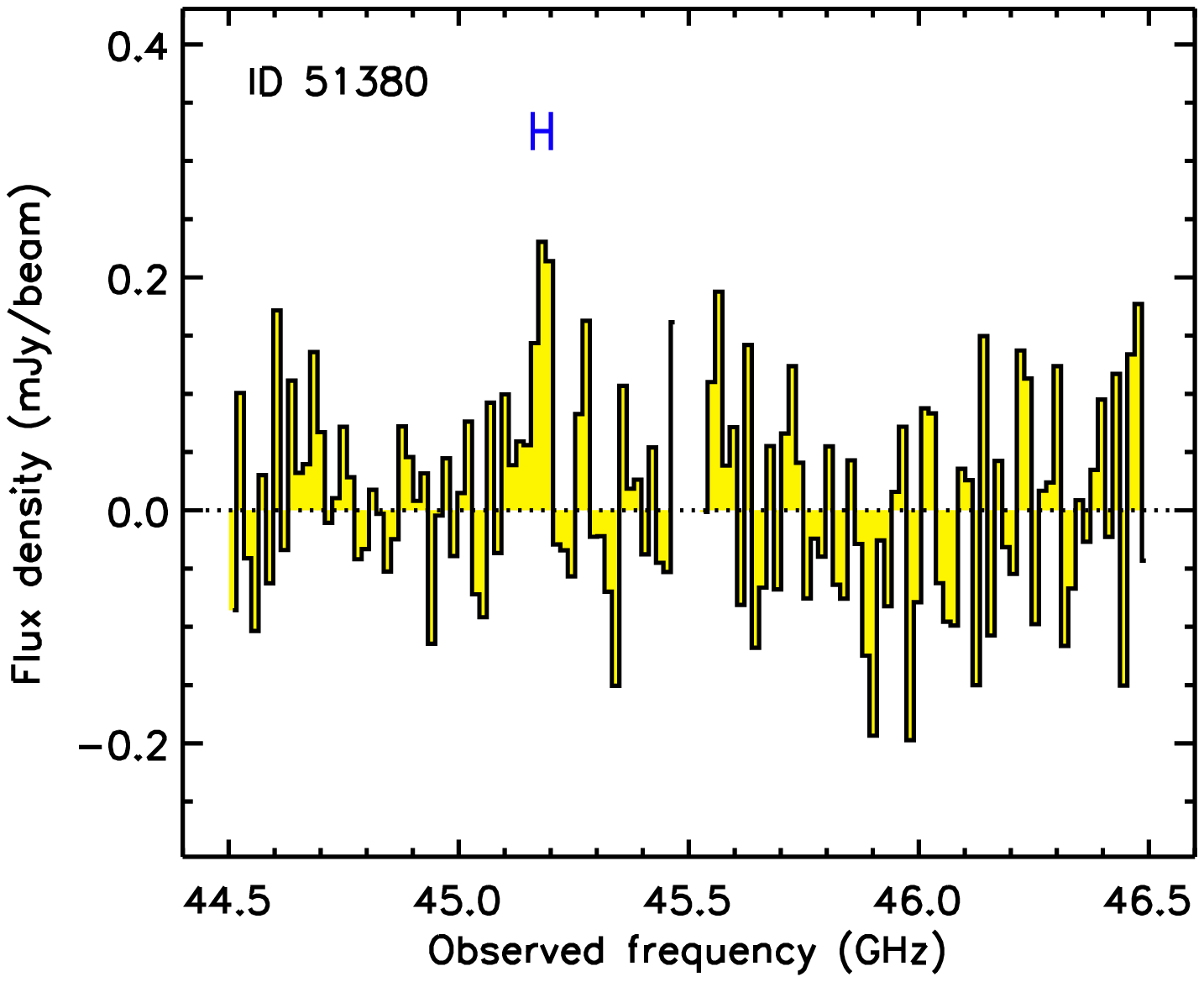}
\caption{JVLA 45.5 GHz spectrum towards ID 51207 and  ID 51380. Channel separation is 16 MHz or $\sim105$ km s$^{-1}$ per channel. The emission profiles were measured at the peak position of the integrated emission. The identified emission feature is shown by a blue horizontal bar with a width equal to the line FWHM. \label{fig:specspec2}}
\end{figure*}

We find that 2 optical sources, ID 51207 and ID 51380, can be identified with emission line peaks from the data cube. The spatial coincidence in both cases is better than $0.6''$. Table \ref{tab:2} summarizes the results and Figs \ref{fig:specmaps2} and \ref{fig:specspec2} show the extracted spectra and images. In both cases, the line peaks have significances $>4\sigma$, however in the case of ID 51207, the positive line peak in the spectrum is followed by a region of noise channels with negative flux density. This is likely because this source is located at the border of the considered JVLA PB FWHM coverage, where the noise starts to increase. 

An additional line association is found for an optical source for which a secure optical spectroscopic redshift of $1.240$ (ID 51861) had been obtained. Its optical redshift discards the possibility of it being a CO detection, since CO $1-0$ could only be detected in the range $\approx1.48-1.59$. At the redshift of this source, no bright emission line from molecules is expected. Therefore, we regard this as a spurious detection. This identification is discussed in more detail in the next sections.

None of the CO sources with optical spectroscopic redshift were found using this method. This is because, at the resolution of the searched data cube ($\sim1''$), ID 51613 has a peak significance of only $\sim3.2\sigma$. In this case, it was necessary to deconvolve the images to a lower resolution of $\sim2''$ to maximize the signal-to-noise. In the case of ID 51858, the significance of the CO detection is too low to be picked up by our signal-to-noise limit.

\subsection{Line search from Bayesian inference}

We performed an independent search for line peaks using model fitting to the data cube upon the principles of Bayesian inference with the MultiNEST algorithm \citep{Feroz2008}. This method provides an efficient way of sampling from the posterior probability distribution and calculating evidence values. We use a galaxy model characterized by a set of parameters $\Theta$ (e.g. x,y positions, line profile central frequency), such that spatially a source is described by a circular or elliptical Gaussian, and in the frequency domain the rotation curve is approximated using either a Gaussian line profile or Brandt's parametrization \citep{Brandt1960}. A full description of this approach to source line searches is given by \citet{Feroz2008}. 

Selection between two models $H_0$ (that there is no source present) and $H_1$ (that there is a source present) can be performed by comparing their evidence values, $Z_1$ and $Z_2$, via $R=Z_1/Z_2$. The probability that there is a source present in the data is then computed from the relation $P=R/(1+R)$.

The four galaxies that have optical spectroscopic redshifts in the range covered by the JVLA observations were searched within a radius of $1''$ of the optical identification, and central frequencies accurate to within 1 channel width ($\sim$100 km s$^{-1}$) from their optical spectroscopic redshifts. No line detection is found for ID 51130 and ID 51480, however we find 55\% and 41\% probability that there is line emission from ID 51613 and ID 51858, respectively. Probability values $>50\%$ are considered as good evidence that there are line detections, given that at this level $<30\%$ of the identified peaks are expected to be spurious line detections. In the case of ID 51858, the lower probability is due to the significant offset between the measured line frequency and the frequency implied by the optical spectroscopic redshift. 

The remaining galaxies with photometric redshifts were searched within a radius of $1''$ from the optical identifications, but central frequencies covering the full bandwidth of the JVLA observations. Using this method, the two identified CO source candidates presented in Section 3.3, ID 51207 and ID 51380, were found to have probabilities of 13\% and 70\% that there is a source present, respectively. ID 51861, however, has a probability of 53\%, even though it is a confirmed spurious detection. Such values must therefore be taken with caution, as they give an estimate of the reliability that the line peak can be described with a physical model of how a CO emission line typically looks like, and they do not take into account the properties of the identified optical sources. The low probability measured for ID 51207 is likely due to the fact that this source lies at a channel, an position (border of the image), where the noise degrades.

Finally, we remark that a total of 7 line candidate detections with $P>50\%$ are found using this method, 6 of which are also found with SERCH. Three of them correspond to sources studied in this work: ID 51380, ID 51613, ID 51861. The rest, despite their high probability, are peaks with lower signal to noise values, as measured directly in the JVLA maps, and therefore not considered further.

\subsection{Probability of chance association}

We estimated the probability that a line peak would be associated by chance within $1.0''$ from an optical galaxy by performing a Monte Carlo simulation of the distribution of significant line peaks in the JVLA field. 

Before starting with simulations, we first carried out a blind search for all positive peaks that have a signal to noise ratio $>4.0$ within the JVLA data cube, using a Gaussian kernel width of 350 km s$^{-1}$. This search yielded a total of 61 line peak candidates within the $60''$ JVLA primary beam. The peaks appear to be slightly clustered in some regions of the image and thus they are not strictly random. From the 61 peaks, 3 correspond to the optical identifications presented in Section 3.3, of which one is spurious as noted in the previous section, and 9 constitute additional line-peak/optical-source associations that do not meet all the specified $K_\mathrm{AB}<23.5$ and $z_\mathrm{phot}=1.2-1.8$ criteria. All these extra associations coincide with optical sources in the redshift range $1.3-2.4$. A total of 49 of the 61 line peaks (80\%)  do not have a clear nearby optical counterpart and likely correspond to spurious detections. This means that up to 20\% of the cases could correspond to real CO line emitting galaxies. We note that we are biasing our counterpart selection by using only the optical $i^+$-band selected source catalog. Thus, line peak detections with no optical counterpart could still represent real sources. For instance, highly obscured sources or at high-redshift ($z>4$) are typically missed by optical catalogs and they can only be revealed in IRAC images \citep[or at far-IR/submillimeter wavelengths; e.g., ][]{Dannerbauer2008}. As mentioned previously, searching for these kind of sources is out of the scope of this paper.

To compute the probability $P$ of chance association between a random peak detection and an optical source within $1''$ radius, we performed Monte Carlo simulations of line peaks in the JVLA region. In each of the $10^4$ samplings, we generated 61 peaks drawn by the real spatial distribution of peaks in the field (we measured this by constructing a density map of peaks using a 7th-nearest neighbor method), and thereby computed $P$ as the fraction of simulated line peaks that are associated to our two optical sources: ID 51207, ID 51380; and also to the spurious identification of ID 51861 for comparison. 

\begin{figure}
\centering
\includegraphics[scale=0.5]{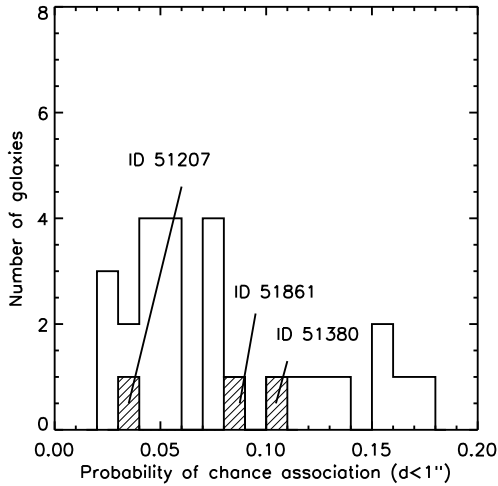}
\caption{Distribution of the probability $P$ that an optical source is associated by chance within $1''$ with any of the 61 significant line peak in the observed region. The dashed histogram shows the value of $P$ for the optical sources found to be associated with emission line peaks. The solid empty histogram shows $P$ for the other optical sources, which met the selection criteria, if they were associated with significant line peaks. Given the low number of significant emission line peaks in the cube, an association by chance is unlikely.\label{fig:probability}}
\end{figure}

Figure \ref{fig:probability} shows the computed probability of chance association for these three sources (filled diagram), as derived from our simulations, compared to the probability that any of the other optical sources that meet our selection criteria would have been associated to a line peak by chance (solid histogram).  For several optical sources that meet our selection criteria, including our line identified targets, the probability to find a line peak within $1''$ is non-negligible, but in all cases less than 20\%. While for ID 51207 the probability is small ($\sim3\%$), in the cases of ID 51380 and ID 51861 there is $\sim10\%$ probability that they are associated by chance. Such probabilities represent an upper limit, since the actual matches in those cases were better than $1''$. This procedure, however, assumes that the optical selection criteria provides good galaxy candidates with redshifts accurate to within $\delta z<0.2$ in a case by case basis. 

In the case of the identification of the optical source ID 51861 with a line peak, it is important to note that even if the photometric redshift estimate ($z_\mathrm{phot}=1.436$) would have agreed better with the spectroscopic one ($z_\mathrm{spec}=1.240$), we would have still selected this source as a possible line candidate given our selection range $z_\mathrm{phot}=1.2-1.8$. However, a critical difference between source ID 51861 with respect to ID 51207 and ID 51380 is that the two later correspond to faint IR detections (see section 4.1). Clearly, detection of IR emission, which is directly linked to dust and gas content, makes a CO detection more likely. The only way to avoid spurious detections is to impose tighter selection criteria, including tighter photometric redshift ranges and detection of sources in the IR bands, at the expense of missing a larger number of emission line candidates.

Finally, we estimated the false match probability by computing the number of associations between negative peaks in the data cube with SNR $<-4\sigma$, searched in the same way than the positive peaks as above, with optical sources that meet our specified selection criteria. From a total of 70 negative peaks, we find only one association within $1''$ from an optical source, and thus imply a $\sim2\%$ false match probability for our positive detections.

\subsection{Dust properties}

\begin{figure*}
\centering
\includegraphics[scale=0.37]{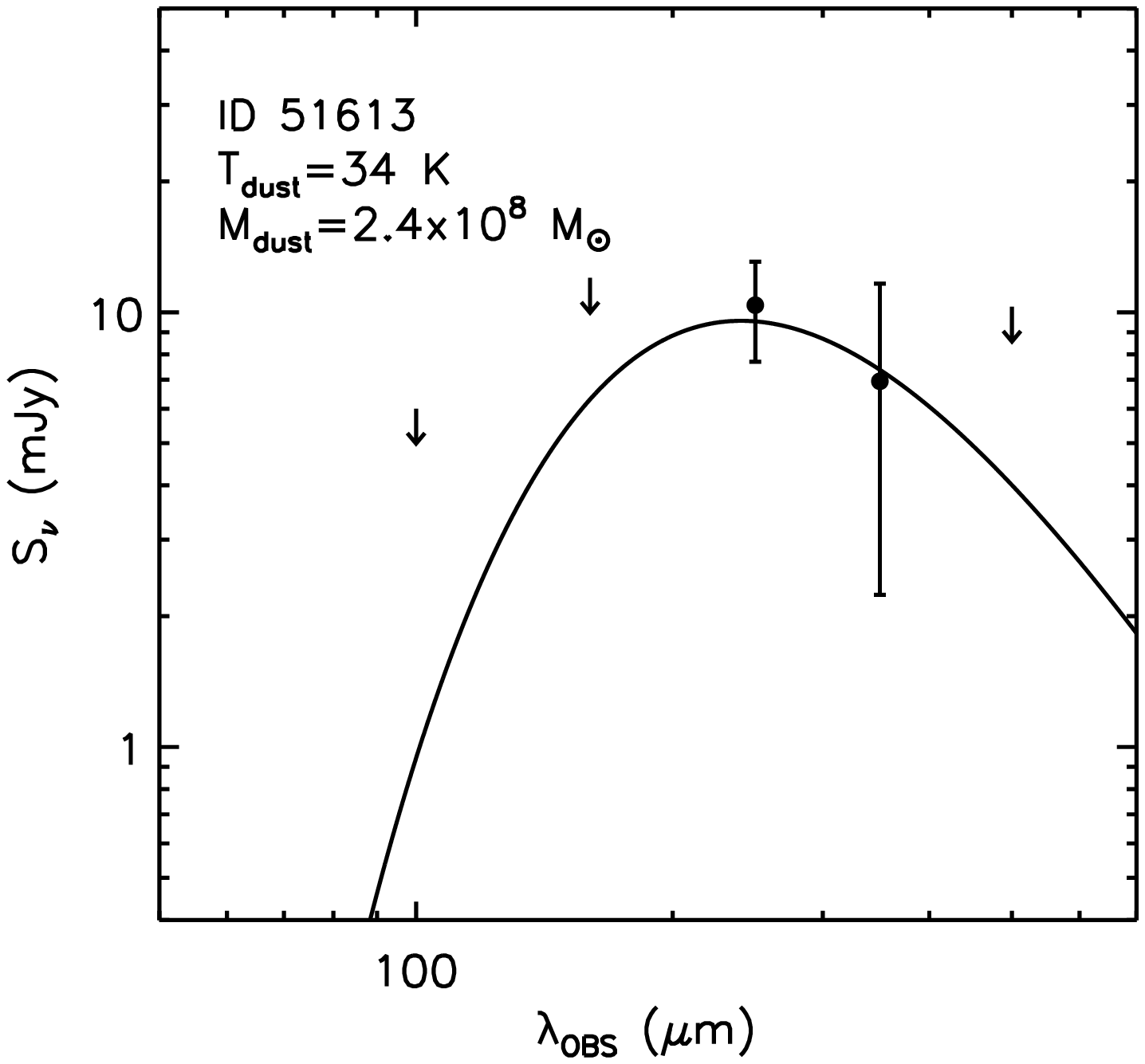}
\includegraphics[scale=0.37]{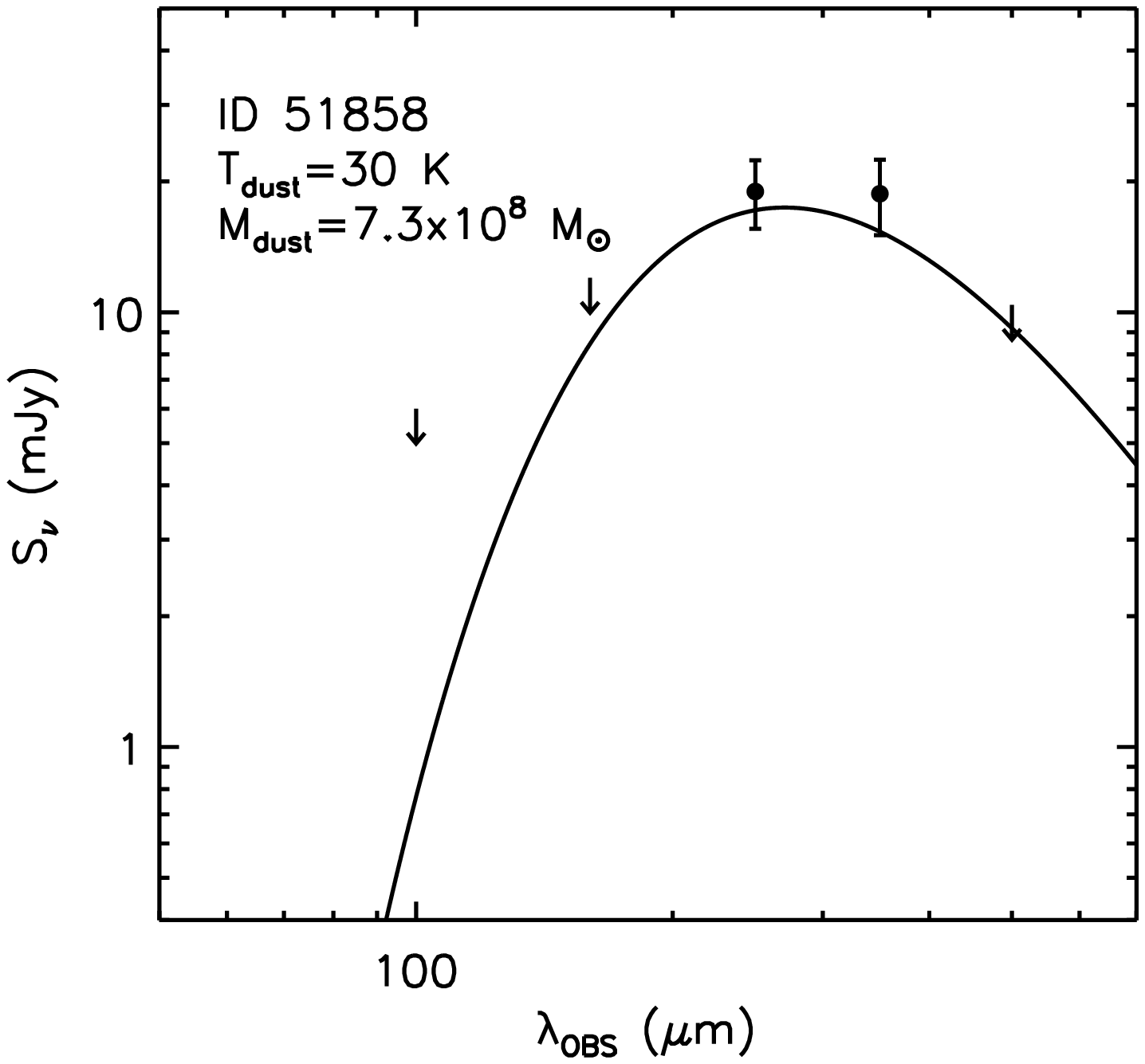}
\includegraphics[scale=0.37]{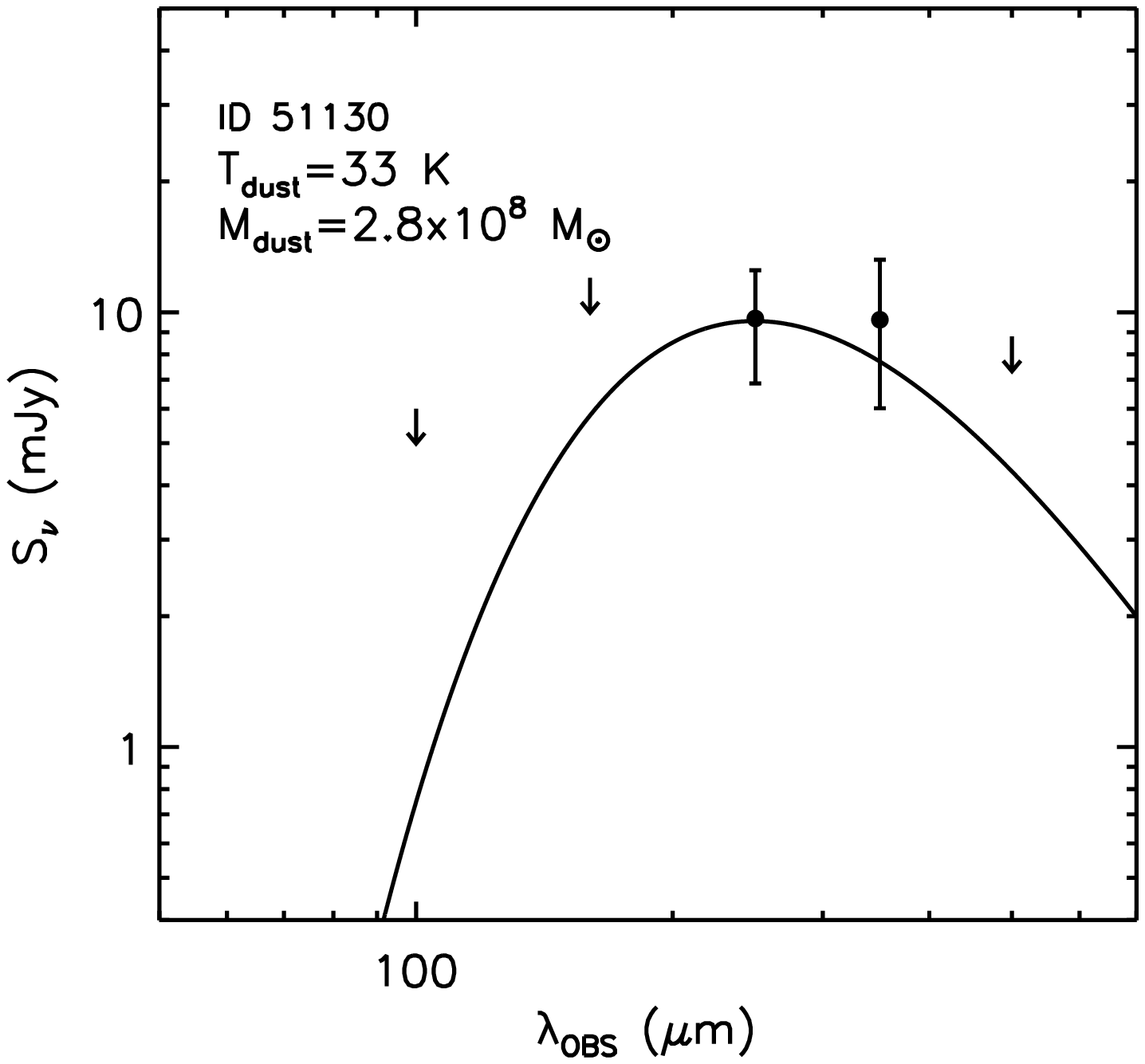}
\includegraphics[scale=0.37]{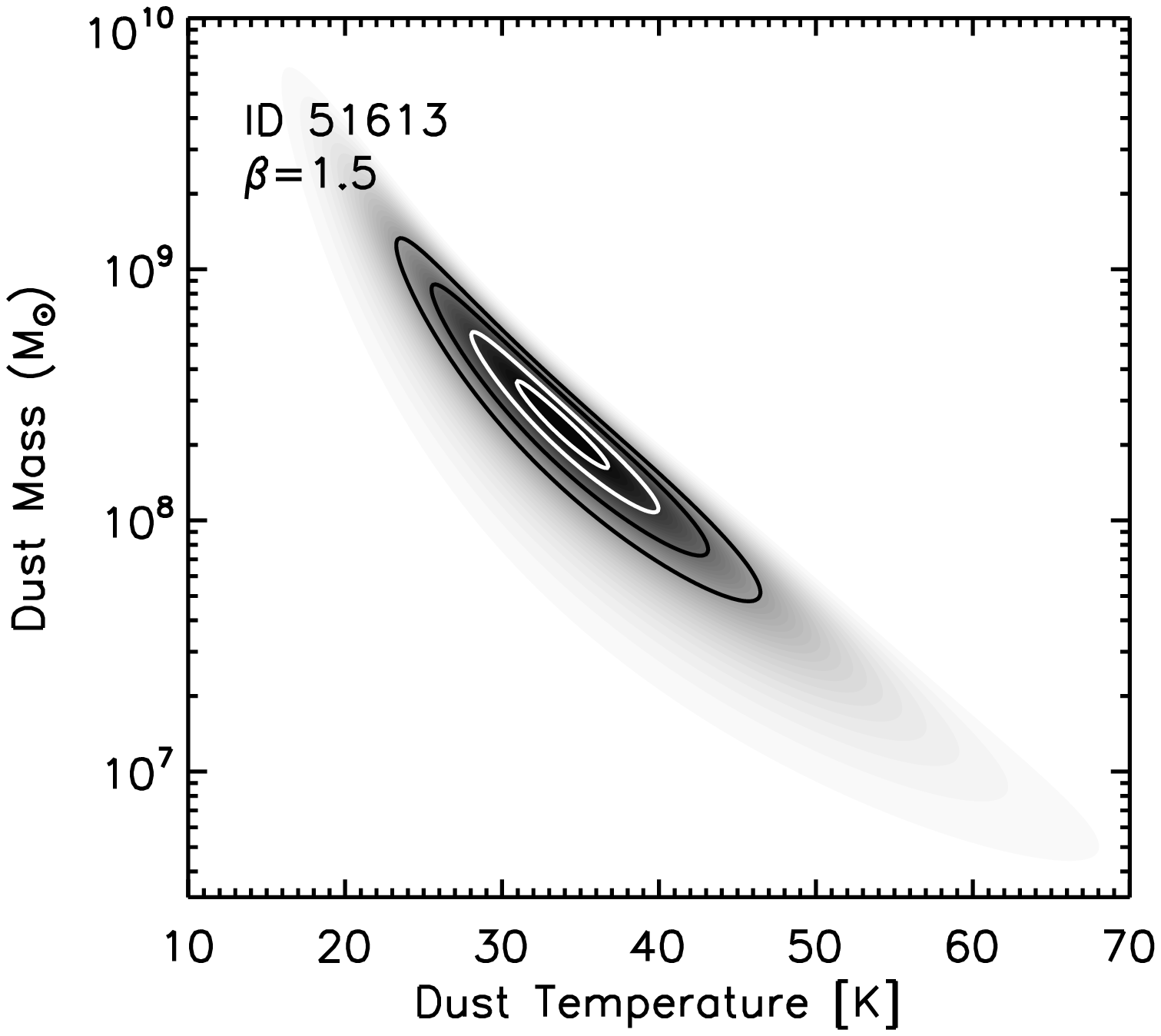}
\includegraphics[scale=0.37]{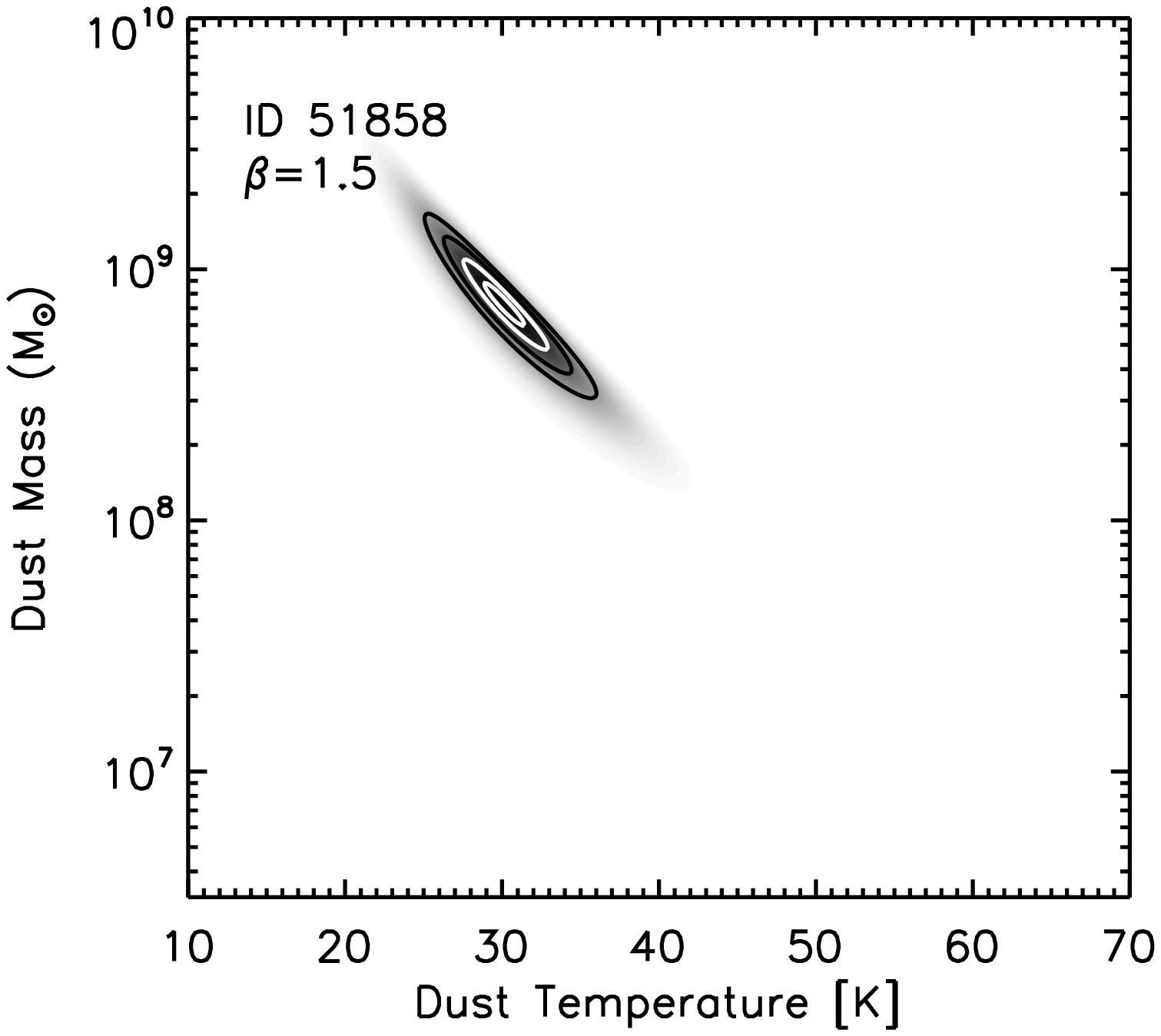}
\includegraphics[scale=0.37]{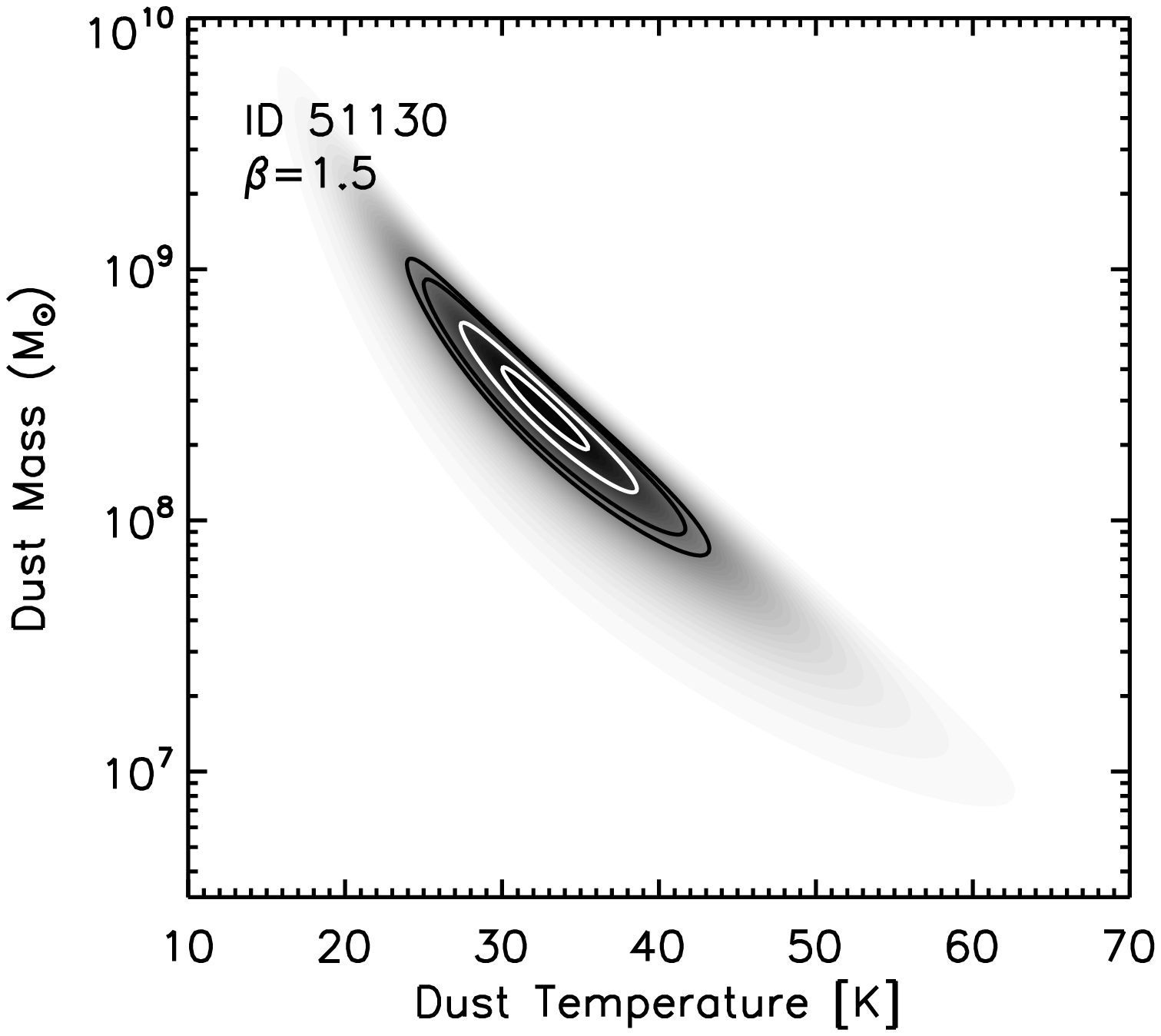}
\caption{{\it Top:} Best fit single-component gray body dust models for objects detected in two bands with {\it Herschel}. {\it Bottom:} Normalized likelihood distribution for different combinations of dust temperatures and masses for $\beta=1.5$, obtained from the comparison between the data and the gray-body models. Contours represent levels at 0.4, 0.6, 0.8 and 0.9.\label{fig:dust}}
\end{figure*}

An important ingredient to characterize the properties of galaxies and to test the reliability of the CO candidate sources is their far-IR emission produced by dust. For the derived CO $1-0$ luminosities, it would be difficult to not detect the sources in the {\it Spitzer} and {\it Herschel} SPIRE maps for typical gas to dust ratios and SFRs. Detection in these IR bands and consistency of the expected IR luminosities can give strong supporting evidence of the reality of the CO line emission.

A number of the galaxies targeted by the JVLA observations have far-IR measurements provided by the {\it Herschel} PACS and SPIRE imaging surveys of the COSMOS field \citep{Lutz2011,Oliver2012}. Since the galaxies in the targeted region are strongly clustered, the emission is blended in the {\it Herschel} images. Fluxes from the PACS/SPIRE maps were thus extracted using the 24 $\mu$m positions as priors, as explained in detail by \citet{Lutz2011}. This provides reliable measurements of the far-IR fluxes, for galaxies detected in the {\it Spitzer} MIPS 24 $\mu$m images.

Within the JVLA field of view (Fig. \ref{fig:field}), we find 10 MIPS 24 $\mu$m sources, 8 of which have photometric redshifts $>1.0$.  Five of these 8 correspond to sources detected with SPIRE at 250 $\mu$m with $z_\mathrm{phot}=1.4-1.6$, 3 of which are also detected at 350 $\mu$m with signal to noise ratio $>4\sigma$. Four of the 5 SPIRE detected galaxies have associated CO detections. The remaining 3 out of the 8 high-redshift 24 $\mu$m sources, are either not detected by {\it Herschel} or lie at $z>2$. None of the 8 high-redshift 24 $\mu$m sources are detected with PACS at 100 or 160 $\mu$m, or with SPIRE at 500 $\mu$m. These data, however, provide important constraints to the far-IR SED. For reference, SPIRE confusion limits on the HerMES fields are 5.8, 6.3 and 6.8 mJy beam$^{-1}$ at 250, 350 and 500 $\mu$m, respectively \citep{Nguyen2010}. In Table \ref{tab:fluxes} we list the measured IR flux densities for the CO detected sources and candidates, including the radio source ID 51130.

For the 3 galaxies that were detected at both 250 and 350 $\mu$m, we fitted a single-component gray-body dust model using the {\it Herschel} photometry. We assume optically thin emission ($\tau_\nu < 1$), with the flux density given by:

\begin{equation}
S_\nu = \frac{(1+z)}{D_\mathrm{L}^2} M_\mathrm{d} \kappa_\nu [B_\nu(T_\mathrm{d}) - B_\nu(T_\mathrm{BG})],
\end{equation}
where $B_\nu(T)$ is the Planck function, and $T_\mathrm{d}$ and $M_\mathrm{d}$ correspond to the dust temperature and mass. $D_\mathrm{L}$ and  $T_\mathrm{BG}$ are the luminosity distance and cosmic microwave background temperature at the source's redshift, respectively. The dust absorption coefficient is parametrized as $\kappa_\nu=\kappa_0 (\nu/\nu_0)^\beta$, where we adopt an emissivity index $\beta=1.5$  \citep{Kovacs2006,Magdis2011} and $\kappa_0=0.4$ cm$^2$ gr$^{-1}$ at 250 GHz \citep{Kruegel1994}. 

Figure \ref{fig:dust} shows the best fitting models to the far-IR data points along with the likelihood distribution, illustrating the uncertainties involved in temperature and mass. We find dust temperatures of $34\pm4$ K, $30\pm2$ K and $33\pm4$ K, and dust masses of $(2.3\pm0.4)\times10^8$ $M_\odot$, $(7.3\pm0.8)\times10^8$ $M_\odot$ and $(2.8\pm0.5)\times10^8$ $M_\odot$ for ID 51613, ID 51858 and ID 51130, respectively. These correspond to total IR luminosities (integrated over the range 8-1000$\mu$m) of $5.7\times10^{11}$ $L_\odot$, $9.5\times10^{11}$ $L_\odot$ and $5.5\times10^{11}$ $L_\odot$, respectively. We note that given the lack of detections at shorter IR wavelengths ($25-70$ $\mu$m rest-frame), it is not possible to estimate the contribution of a hot dust component, although the contribution to the IR luminosity is likely negligible. 

In the two cases where only the 250 $\mu$m emission was detected, we fit only the dust mass (and hence the IR luminosity) by fixing the dust temperature to $35$ K, which is the typical value found for star-forming galaxies at high-redshift \citep{Kovacs2006,Magdis2011}. We find dust masses of $1.7\times10^8$ $M_\odot$ and $1.1\times10^8$ $M_\odot$, and IR luminosities of $4.9\times10^{11}$ $L_\odot$ and $3.3\times10^{11}$ $L_\odot$ for ID 51207 and ID 51380, respectively. Varying the assumed dust temperature in a typical range $30-40$ K, leads to dust masses in the range $(1.0-3.4)\times10^8\ M_\odot$ and $(0.7-2.3)\times10^8\ M_\odot$, and thus IR luminosities in the range $(4.2-6.0)\times10^{11}\ L_\odot$ and  $(2.8-4.0)\times10^{11}\ L_\odot$, for ID 51207 and ID 51380, respectively. This corresponds to $\sim30\%$ uncertainty in luminosity caused by our assumption in dust temperature. Finally, we remark that given the lack of sufficient constraints on the far-IR SED for these galaxies, our luminosity estimates could be affected by different choices of $\beta$, as well as dust temperatures, and thus should be taken with caution. 

\begin{table*}
\centering
\caption{Physical properties \label{tab:photo}}
\begin{tabular}{cccccccccccc}
\hline
ID $^a$ & $z_\mathrm{spec}$ $^b$ & SFH $\tau$ $^c$& Age $^d$& $M_\mathrm{V}$ $^e$ & $A_\mathrm{V}$ $^f $ & $M_\mathrm{stars}$ $^g$ & SFR$_\mathrm{SED}$ $^h$ & SFR$_\mathrm{IR}$ $^i$ & SFE $^j$\\
     &                                   & (Gyr)            & (Gyr) &        (mag)       &    (mag)       & ($\times10^{10}\ M_\odot$) & ($M_\odot$ yr$^{-1}$) & ($M_\odot$ yr$^{-1}$) &($M_\odot$ yr$^{-1}$) & $l_0$\\   
\hline \hline
50480 & 1.523 & 0.1  & 0.25 & -21.1 & 0.2  & $3.5_{-0.2}^{+0.2}$ & 4 & $\ldots$ & $>4$ \\
51130 & 1.519 & 0.3  & 0.50 & -24.0 & 1.3  & $24_{-2}^{+4.0}$  & 25 & 94 & $>80$ \\
51613 & 1.516 & 0.6  & 0.36 & -22.4 & 1.9  & $4.5_{-1.5}^{+1.8}$ & 130 & 97 & 23\\ 
51858 & 1.556 & 3.0  & 0.36 & -22.7 & 2.0  & $6.0_{-1.0}^{+1.6}$ & 218 &162 & 75\\
\hline
51207 & 1.530 & 0.6  & 1.00 & -21.9 & 2.5  & $11_{-1.3}^{+5.7}$   &  66 & 83 & 47 \\
51380 & 1.551 & 0.6  & 0.25 & -21.4 & 2.6  & $ 2.8_{-1.1}^{+1.3}$ &  120 & 56 & 29\\
\hline
\end{tabular}
\begin{flushleft}
\begin{footnotesize}
\noindent $^a$ COSMOS ID; $^b$ CO-based spectroscopic redshift or optical, if CO undetected ; $^c$ Timescale of exponentially declining star-formation history (SFH) best-fit template. $^d$ Age of the best-fit template SED; $^e$ Optical absolute magnitude (AB); $^f$ Best-fit extinction correction; $^g$ Best-fit stellar mass. Uncertainties correspond to the 90\% confidence interval; $^h$ Star formation rate (SFR) derived from SED fitting; $^i$ Star formation rate derived from IR luminosity, assuming SFR/($M_\odot$ yr$^{-1}$)$=1.7\times10^{-10}\ L_\mathrm{IR}/L_\odot$; $^j$ Star formation efficiency, SFE$=L_\mathrm{IR}/L'_\mathrm{CO}$ in units of $l_0=L_\odot$ (K km s$^{-1}$ pc$^2$)$^{-1}$.
\end{footnotesize}
\end{flushleft}

\end{table*}

\begin{table}
\centering
\caption{Flux densities measured with {\it Spitzer} and {\it Herschel}\label{tab:fluxes}}
\begin{tabular}{ccccc}
\hline
ID$^{a,b}$ & $S_{24}\ ^c$ &  $S_{250}\ ^{d}$& $S_{350}\ ^{d}$& $S_{500}\ ^{d}$\\
    &  ($\mu$Jy) & (mJy) & (mJy) & (mJy)\\
\hline\hline
51130 & $220\pm30$ & $9.7\pm1.7(2.8)$  &  $9.6\pm2.1(3.6)$ & $<8.8$ \\
51613 & $125\pm20$ & $10.4\pm1.5(2.7)$ & $6.9\pm3.7(4.7)$ & $<10.3$ \\
51858 & $150\pm15$ & $19.0\pm2.6(3.4)$ &  $18.7\pm2.3(3.7)$ & $<10.4$ \\
51207 & $110\pm15$ & $7.7\pm1.5(2.7)$   &  $<9.6$  & $<7.7$ \\
51380 & $65\pm17$ & $5.0\pm0.5(2.3)$    & $<13.3$ & $<9.7$ \\
\hline
\end{tabular}
\begin{flushleft}
\begin{footnotesize}
\noindent $^{a}$COSMOS ID; $^b${\it Herschel} PACS measurements led to $3\sigma$ upper limits of 6 and 12 mJy at 100 and 160 $\mu$m, respectively; $^c$Flux density measured with the {\it Spitzer} MIPS at 24$\mu$m; $^d$Flux densities measured with {\it Herschel} SPIRE at 250, 350 and 500 $\mu$m and were computed using the 24$\mu$m position as prior \citep{Lutz2011}. Quoted uncertainties correspond to the map flux error and total uncertainty (including instrumental errors), shown in parenthesis, at the given band.
\end{footnotesize}
\end{flushleft}
\end{table}

\subsection{Galaxy properties}

We use the full multi-wavelength dataset available in the COSMOS field, including 31 broad, intermediate and narrow bands from the UV/optical and the IRAC 3.6$\mu$m, 4.5$\mu$m, 5.8$\mu$m and 8.0$\mu$m bands, to measure the range of physical properties of the observed galaxies in the JVLA field. 

We computed SFRs and stellar masses using the SED fitting routine HyperZ \citep{Bolzonella2000} and fixing the spectroscopic redshift as given by the optical spectra or the CO line identification, under the assumption that the latter is correct for the two CO identified candidates (ID 51207 and ID 51380). We made use of an updated version of the COSMOS photometric catalog \citep[see][]{Salvato2011}; and applied the appropriate galactic extinction corrections, and ``auto-offsets'' for SED fitting computation \citep{Ilbert2009, Salvato2011}.

\begin{figure*}
\centering
\includegraphics[scale=0.5]{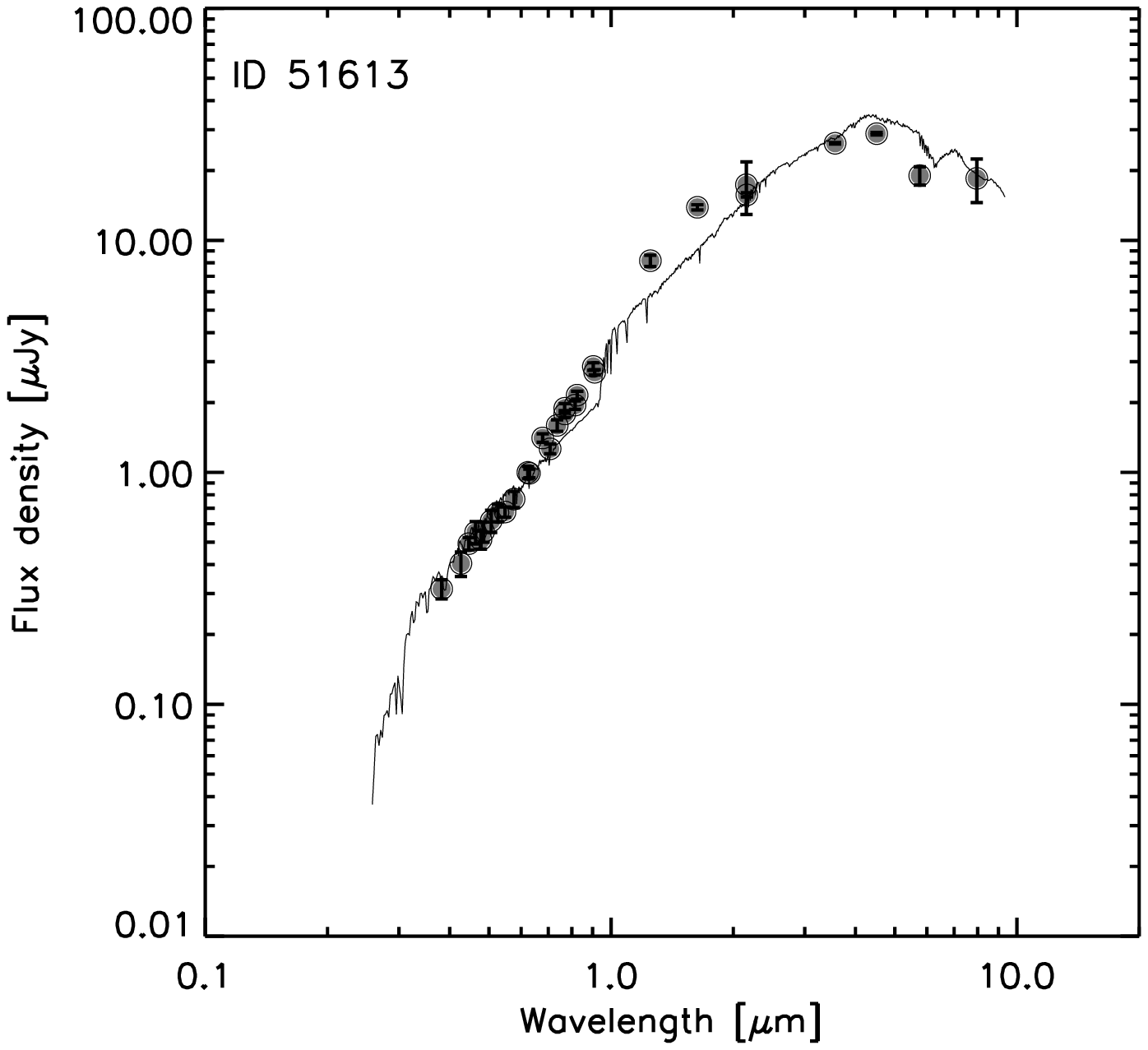}
\includegraphics[scale=0.5]{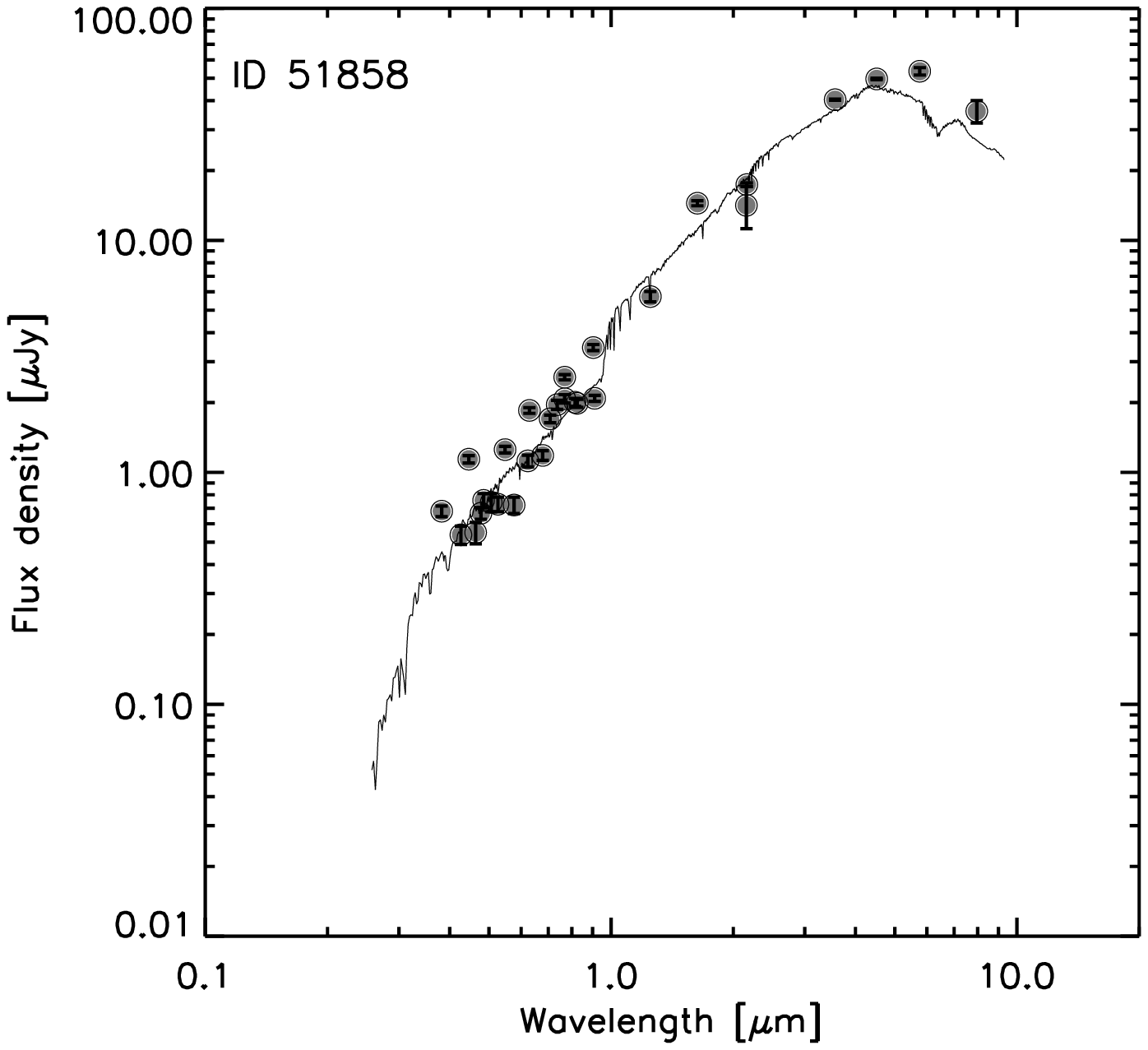}
\includegraphics[scale=0.5]{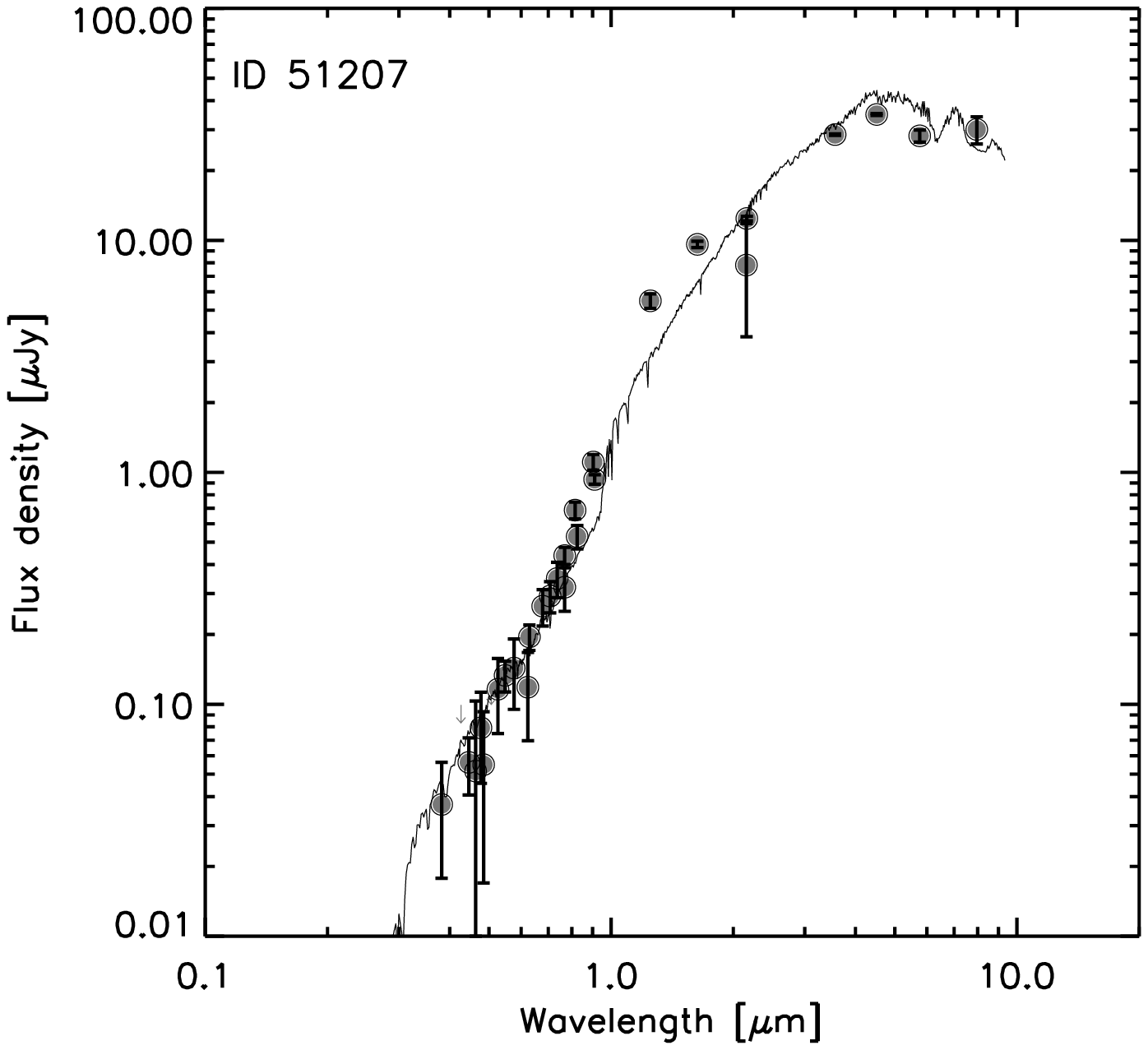}
\includegraphics[scale=0.5]{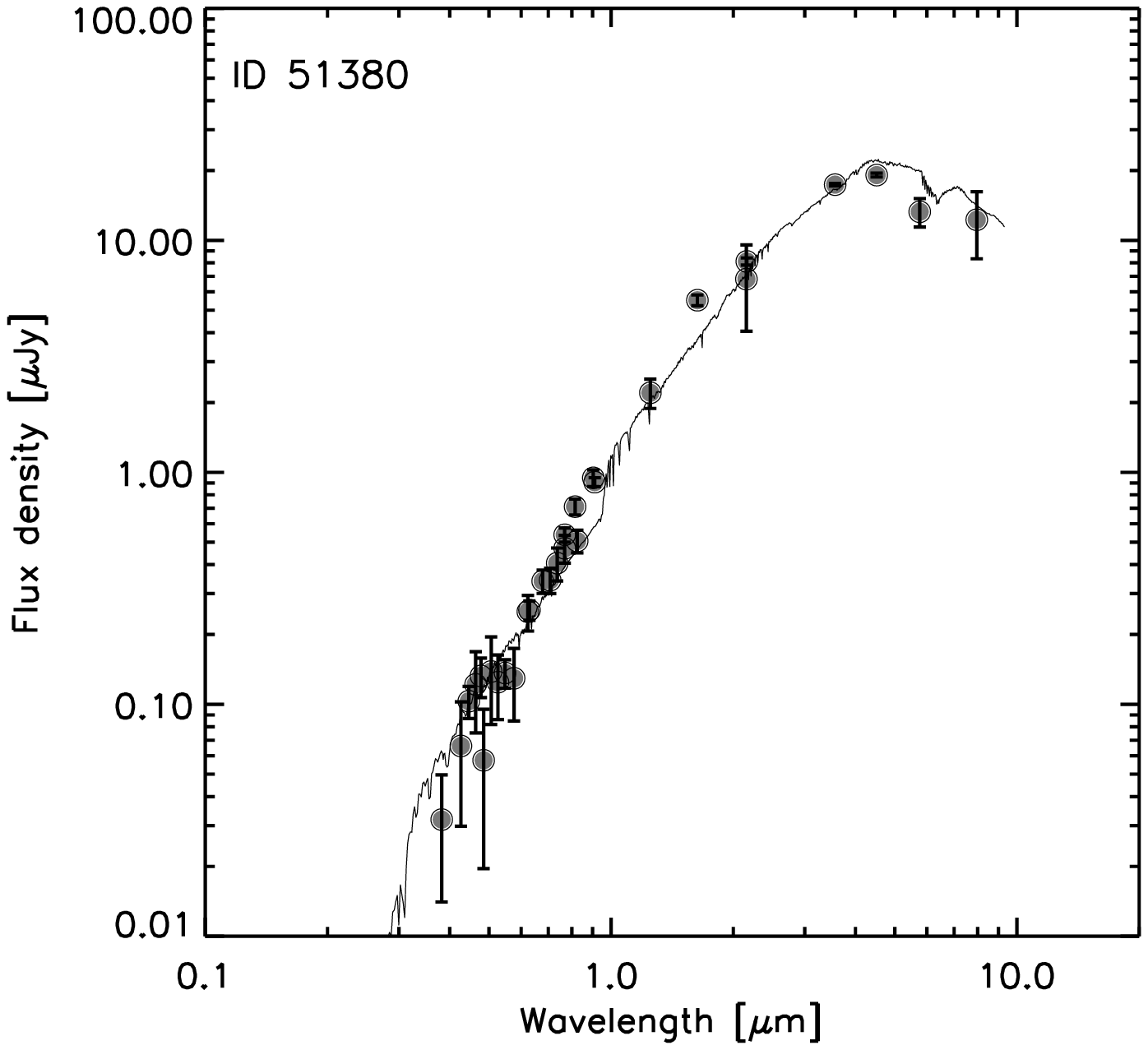}
\caption{Optical to mid-IR spectral energy distribution (SED) toward both CO detected galaxies (ID 51613 and ID 51858) and CO candidates (ID 51207 and ID 51380). Data points are drawn from the COSMOS field multi-wavelength photometry. Solid lines show the best fitting templates from the libraries of \citet{Bruzual2003}. \label{fig:sed}}
\end{figure*}

We utilized 6 synthetic template sets from the \citet{Bruzual2003} libraries, including a constant star formation history (SFH) and 4 exponentially declining ones, with typical decaying timescales of 0.1, 0.3, 0.6, 3.0 and 10 Gyr. We used an extinction correction from \citet{Calzetti2000}, with an optical extinction $A_\mathrm{V}=0-3$. Both the best fit template and the fixed redshift give the absolute magnitude, which combined with a prescription for the mass-to-light ratio and SFR from the UV/optical continuum for each template, can be used to derive the stellar masses and SFR for each galaxy. Due to degeneracies in the extinction correction and best template fit, and the use of templates that do not include emission lines, these estimates lead to typical uncertainties of $30-50\%$, similar to what was reported in other similar studies \citep{Daddi2010a, Tacconi2010}. Table \ref{tab:photo} lists the obtained parameters for the best fitting template SEDs. For the CO detected and CO candidate galaxies, we also show the best fitting template at the derived CO redshift (Fig. \ref{fig:sed}).

From the good agreement between the spectroscopic and the photometric redshifts for the galaxies in our field, we use the results from \citet{Ilbert2009} as a cross-check for the parameters used in our fitting routine (e.g. template libraries, extinction). Such measurements provide an independent estimate of the stellar masses and SFRs at the best fit photometric redshift. We find that while both SFR estimates are in agreement with a typical uncertainty of $30\%$, the stellar mass estimates are in agreement to within 15\%. The larger difference between both SFR estimates, compared to the difference in stellar masses, reflect the uncertainties in the extinction correction applied. While estimates for the stellar masses largely rely on the rest-frame optical and near-IR emission, which is less affected by extinction, estimates for the SFR are based on the rest-frame UV for which a proper extinction correction factor plays a key role.

An independent and more direct measure for the SFR is given by the far-IR emission, and we use it as the actual SFR in the following of the text. From \citet{Kennicutt1998}, we have $\mathrm{SFR}/[M_\odot yr^{-1}]=1.7\times10^{-10} L_\mathrm{IR}/L_\odot$. The IR-derived and SED fitting derived estimates are in agreement to within 25\%, and have a dispersion of only $\sim10\%$ from the average SFR (see Table \ref{tab:photo}). The only exception is ID 51130, where the SED fitting routine implies SFR$=25\ M_\odot$ yr$^{-1}$, while the IR luminosity indicates SFR$=93.5\ M_\odot$ yr$^{-1}$. This difference could be explained by the fitted template SED that does not represent the real nature of this source, despite it provides a good fit to the photometric data points (low $\chi^2$). This large difference may also reflect the uncertainties in the derived IR luminosity given that the IR data is close to the SPIRE confusion limit. Table \ref{tab:photo} lists the computed SFR values.

\section{Discussion}

\subsection{Star-formation rates and efficiencies}

\begin{figure*}
\centering
\includegraphics[scale=0.8]{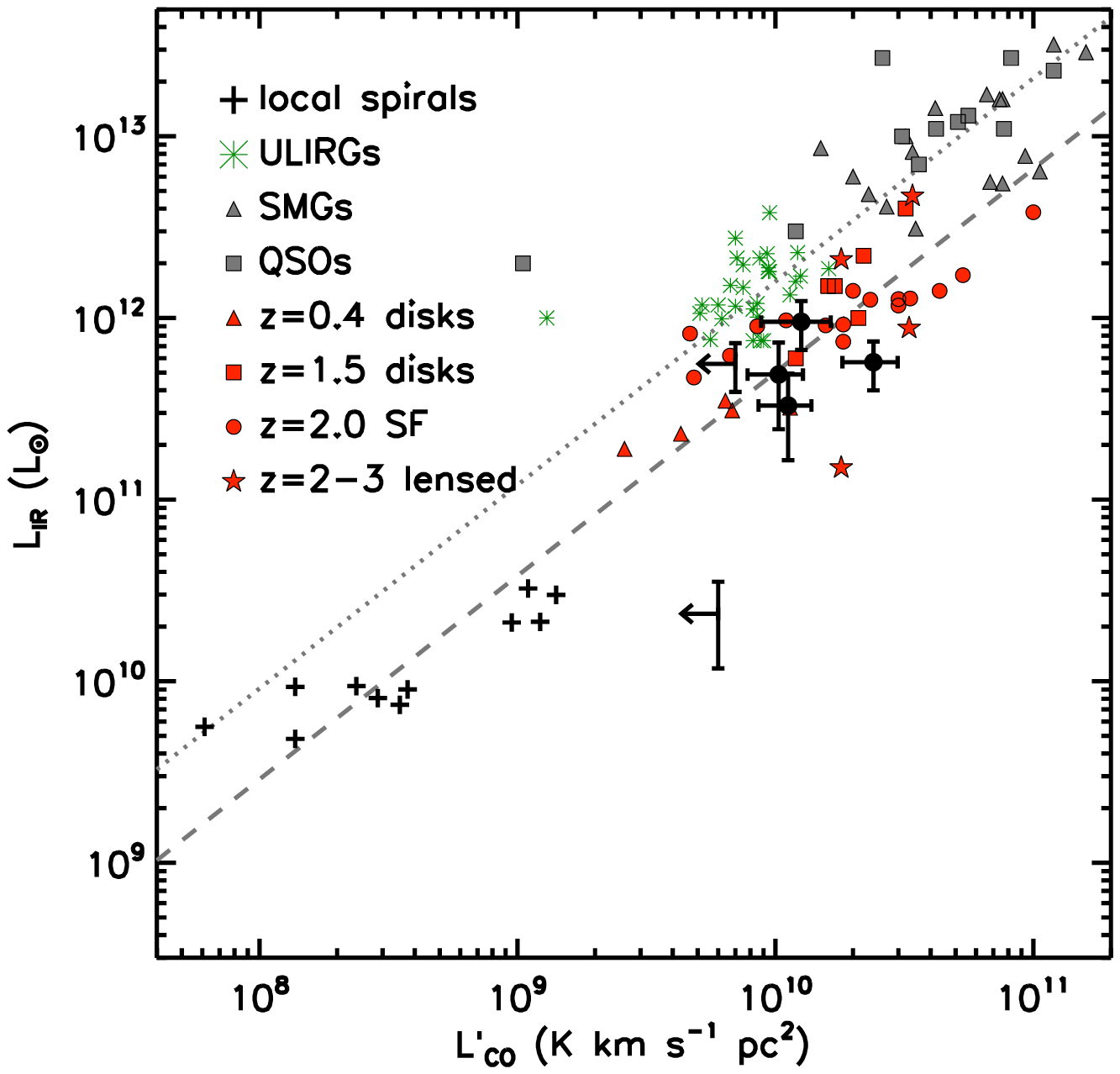}
\caption{CO $J=1-0$ and IR luminosities of the CO-detected galaxies in our cluster field, shown by black filled circles, compared with other galaxy samples. Black crosses show measurements for local spiral galaxies \citep{Leroy2008} and green asterisks represent local ULIRGs \citep{Solomon1997}. Grey triangles and squares show high-redshift SMGs and QSO host galaxies, respectively (see text). Red triangles, squares and circles show massive disk galaxies at $z\sim0.4$ \citep{Geach2011}, $z\sim1.5$ \citep{Daddi2010a}, and typical massive star forming galaxies at $z\sim1-2$ \citep{Tacconi2010}, respectively. Red 5-pointed stars show four lensed massive star-forming galaxies at $z=2-3$, which have intrinsic properties similar to $z\sim1-2$ disks \citep{Riechers2010b, Frayer2011}. In cases were only $J>1$ CO lines are available, we corrected to the CO $J=1-0$ line (see text). Following \citet{Daddi2010a}, we show a representative fit to local spirals and disk galaxies at high-redshift (grey dashed line), log$_{10}\ L_\mathrm{IR}=1.12\times$ log$_{10} L_\mathrm{CO}+0.5$, for guidance. The dotted line shows the same line, with a factor $+0.5$ added.\label{fig:sfe}}
\end{figure*}

Recent studies suggest that, compared to starbursting ULIRGs and SMGs, spiral and disk galaxies have substantial differences in SFEs and dynamical timescales \citep{Daddi2010a, Daddi2010b, Genzel2010}. This has been interpreted as evidence for different modes of star-formation. Disks appear to evolve secularly, with long dynamical and gas consumption timescales, while starbursts seem to evolve rapidly, possibly triggered by merger-induced activity. However, it has been argued that such fundamental differences in the SFEs are, at least partially, produced by observational reasons. Aside from the large uncertainties in CO and IR measurements in ultra-luminous IR galaxies (ULIRGs) and SMGs, the large majority of the CO observations have been done in high-J CO transitions ($J>2$). Extrapolation from high-J CO luminosities to CO $1-0$, a direct tracer of molecular gas mass, is not straightforward in a case by case basis, as CO $1-0$ measurements in a few SMGs found $\sim2\times$ more molecular gas than predicted by the CO $3-2$ emission line \citep{Bothwell2010, Ivison2011, Riechers2011a}. Furthermore, such differences in SFE are increased when using different CO luminosity to gas mass conversion factors for starburst and disks. Until accurate measurements of the CO $1-0$ emission line over a statistical number of galaxies (both disks and SMGs), it will be difficult to solve this issue. 

Although it is out of the scope of this paper to study these important differences, it is instructive to compare our CO $1-0$ measurements with other galaxy populations. Figure \ref{fig:sfe} compares the CO and IR luminosities of different galaxy populations including our detections in the cluster field. We do not convert CO luminosities to molecular gas masses given the large potential uncertainties related to $\alpha_\mathrm{CO}$ conversion factor across different galaxy populations. For the local Universe, we compiled measurements of typical spiral galaxies \citep{Leroy2008} and local ULIRGs \citep{Solomon1997}. Recent CO measurements of massive star-forming galaxies at $z\sim0.4$ \citep{Geach2011}, $z\sim1.5$ \citep{Daddi2010a, Aravena2010} and $z=1-2$ \citep{Tacconi2010} are also shown, as well as CO measurements of four lensed star-forming galaxies at $z=2-3$ that have intrinsic properties similar to $z\sim1-2$ disks \citep{Riechers2010b, Frayer2011}. Additionally, we have included observations of SMGs and IR luminous QSO host galaxies from the literature \citep{Greve2005, Solomon2005, Iono2006, Weiss2007, Aravena2008, Coppin2008, Frayer2008, Schinnerer2008,Ivison2011, Coppin2010, Carilli2010, Carilli2011, Riechers2006, Riechers2010a,Riechers2011a, Riechers2011b, Riechers2011c, Fu2012}\footnote{In the case of the \citet{Greve2005} SMG objects, we recomputed the IR luminosities by considering a optically thin, single component gray-body dust model of the form $S_\nu\propto\nu^\beta B_\nu(T_d) M_d$, with a dust temperature of 37 K  and $\beta=1.5$, typical of star-forming SMGs \citep{Kovacs2006}, constrained by the source's redshift and 350 GHz flux densities. This implies IR luminosities $\sim2\times$ lower than reported by \citet{Greve2005}. For detections made only in a few high-J CO transitions ($J>1$), with no CO excitation models available, we adopt line temperature ratios: $R_{31}\sim R_{41}\sim0.5$ and $R_{21}\sim0.8$ \citep{Ivison2011, Riechers2011a}.}

The SFE can be defined as SFE$=L_\mathrm{IR}/L'_\mathrm{CO}$ in units of $L_\odot $ (K km s$^{-1}$ pc$^2$)$^{-1}$. For galaxies detected with {\it Herschel}, we directly use their estimate of the IR luminosity. For those galaxies with no {\it Herschel} detection, we adopt the SED fitting derived SFR and convert them into IR luminosities following the relation of \citet{Kennicutt1998}. The obtained SFE values are shown in Table \ref{tab:photo}. For the CO detected galaxies, the SFE ranges between 20 to 75 $L_\odot $ (K km s$^{-1}$ pc$^2$)$^{-1}$. 

Figure \ref{fig:sfe} shows that the SFE from the galaxies in our sample are overall consistent with the general population of star-forming galaxies at high-redshifts, which range between 30 to 180 $L_\odot $ (K km s$^{-1}$ pc$^2$)$^{-1}$ \citep{Daddi2010a, Aravena2010, Tacconi2010}, however, the significant uncertainties in our measurements make it difficult to draw further conclusions. On the other hand, the agreement between the SFE of ID 51207 and ID 51380 with typical high-redshift galaxies, gives fundamental support to the identification of those objects as CO line emitters. 

As mentioned before, the optical source ID 51861 has an optical spectroscopic redshift of $1.240$ although it is wrongly identified with a line peak. In the absence of a spectroscopic redshift, it would be identified with CO $1-0$ at $z=1.497$. This source is not detected at 24$\mu$m and hence, by definition, not detected with {\it Herschel} since we are using the 24 $\mu$m prior {\it Herschel} catalog. It is interesting to note that if we had identified this galaxy as a CO emitter at the respective redshift, the SFR obtained from SED fitting of 10 M$_\odot$ yr$^{-1}$ would have implied a SFE of $\sim4\  L_\odot $ (K km s$^{-1}$ pc$^2$)$^{-1}$. This would have been lower than the ones measured in local spiral galaxies, which are typically in the range $10-100$ $L_\odot $ (K km s$^{-1}$ pc$^2$)$^{-1}$ \citep{Leroy2008}, yet still consistent with that found in local elliptical galaxies, in the range $\sim0-50$ $L_\odot $ (K km s$^{-1}$ pc$^2$)$^{-1}$ \citep{Shapiro2010, Crocker2011}. If we did not have an optical spectroscopic determination of the redshift, such mis-identification would have still been suspicious taking into account such low SFE and the overall multi-wavelength properties of the host galaxy. 
\begin{figure}
\centering
\includegraphics[scale=0.55]{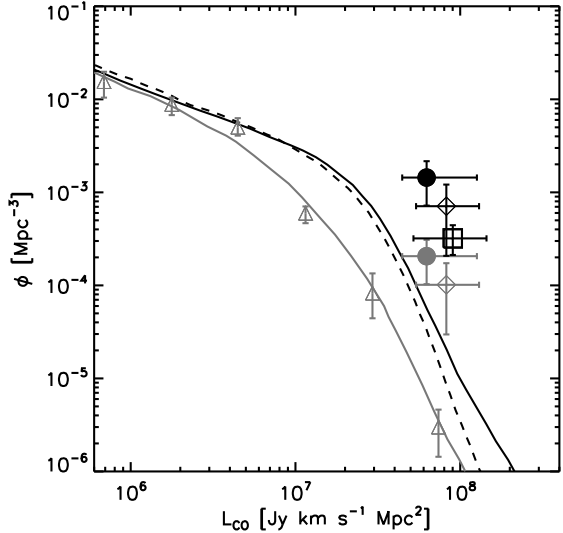}
\caption{Predicted cosmic evolution of the space density of CO $1-0$ emitting galaxies as a function of CO $1-0$ luminosity (CO luminosity function) at redshifts $z=0$ (gray solid line), $z=1$ (black dashed line) and $z=2$ \citep[black solid line; ]{Obreschkow2009b}. The gray open triangles show the space density of CO emitters measured in the local Universe \citep{Keres2003}. The black filled circle shows the space density derived from our 4 CO detections in the overdense field targeted with JVLA. The gray filled circle shows the measurement but decreased by $7$ to correct for the overdensity. The black and gray open diamonds represent the space density obtained considering only the 2 spectroscopically confirmed galaxies uncorrected and corrected by the $7\times$ factor, respectively. The black open square shows the estimate of space density of CO emitters derived from the 6 $BzK$ star-forming galaxies at $z\sim1.5$ \citep{Daddi2010a}. \label{fig:lumfunction}}
\end{figure}

\subsection{Comparison to the predicted CO luminosity function at $z=2$}

One of the major motivations to perform blind deep CO surveys of the sky is to measure possible evolutionary trends in the amount of molecular gas with redshift and CO luminosity (or gas mass). An important tool to do this is to estimate the CO luminosity function at different redshifts. Combining results from the Millennium simulation \citep{Springel2005, DeLucia2007} and empirical, detailed models of abundances of molecular gas, and their relation to CO intensities,  \citet{Obreschkow2009b}, made predictions for the luminosity function of the CO emission lines as a function of redshift and CO rotational transitions. In this section, we measure the space density of CO detected galaxies within the physical volume covered by our observations, and compare this with the values expected from the predictions \citep{Obreschkow2009b} and with previous observations of massive star-forming galaxies at $z\sim1.5$ \citep{Daddi2010a}.  

To estimate the space density implied by our 4 CO detections, we start by computing the CO luminosity in units of Jy km s$^{-1}$ Mpc$^2$ as $L^\mathrm{V}=(1+z)^{-1} 4 \pi D_\mathrm{L}^2 S^\mathrm{V}$ \citep[adopting the terminology presented in the appendix A of ][]{Obreschkow2009b} , where $z$ is the redshift, $D_\mathrm{L}$ is the luminosity distance in Mpc, and $S^V$ is the velocity-integrated CO intensity in units of Jy km s$^{-1}$. From our 4 detections, we estimate a weighted average CO luminosity of $(6.3\pm1.6)\times10^7$ Jy km s$^{-1}$ Mpc$^2$. Next, we determine the cosmological volume covered by our observations, considering the area comprised by our JVLA field of view ($\approx60''\times60''$) and the redshift coverage, $z=1.478-1.591$. This leads to a volume of $\Delta V=356$ Mpc$^{3}$. With this, we compute the space density of CO sources in units of Mpc$^{-3}$ dex$^{-1}$ as $\phi_0 = N/(\Delta V \mathrm{log}_{10} L_\mathrm{CO})$, where $N=4$ is the number of CO detected galaxies (Fig. \ref{fig:lumfunction}).

We also compute the space density of CO emitters implied by the sample of 6 massive star-forming BzK galaxies that were reliably detected in CO $2-1$ line emission by \citet{Daddi2010a}.  These are typical massive galaxies with an average IR luminosity of $\sim2\times10^{12}\ L_\odot$. For this, we follow the procedure used by \citet{Obreschkow2009a}, but we adopt more recent developments to our analysis, including the use of 6 BzK galaxies available instead of only 2 from \citet{Daddi2008}. For each object, we compute its CO $2-1$ luminosity directly from its redshift and measured CO line intensity presented by \citet{Daddi2010a}.  We convert the CO $2-1$ luminosities (in Jy km s$^{-1}$ Mpc$^2$) into CO $1-0$ luminosities by assuming that both transitions are typically in local thermodynamic equilibrium (LTE) in BzK galaxies, and therefore scaling approximately as $\nu^2$ \citep{Aravena2010}. Thereby, we compute the weighted average CO $1-0$ luminosity for the sample. Following \citet{Obreschkow2009a}, we approximate the space density of CO sources as the space density of IR sources at $L_\mathrm{IR}\approx2\times10^{12}\ L_\odot$, which corresponds to the average IR luminosity for all 6 objects. For typical galaxies at this IR luminosity level and in the redshift range $z=1.2-2.0$, \citet{Rodighiero2010} finds a space density $\Phi_\mathrm{IR}=3.2_{-0.5}^{+0.6}\times10^{-4}$ Mpc$^{-3}$ per unit log$_{10} L_\mathrm{IR}$, which we scale to log$_{10} L_\mathrm{CO}$ to find the space density of CO sources (Fig. \ref{fig:lumfunction}). We selected only this sample of massive star-forming galaxies at high redshift, with respect to the sample from \citet{Tacconi2010}, because (1) all of them were detected in CO $2-1$ with high significance; (2) they have similar redshifts compared to our CO sources; and (3) three of these sources correspond to the only `normal' objects so far detected in CO $1-0$ emission, making a determination of line ratios between the $2-1$ and $1-0$ transitions more direct and straightforward. We remark, however, that this method to estimate the space density of CO emitters is limited by two important assumptions: the main one that the space density of CO emitters is directly traced by the space density of IR sources at a given IR luminosity does not take into account that there appears to be different star formation laws for bright starbursts and the general population of galaxies \citep[see previous section;]{Daddi2010b}. Without knowledge of the nature of individual sources in the IR sample from where we adopt the space density, it is difficult to estimate the uncertainties introduced by this assumption. Furthermore, the 6 BzK galaxies were selected to be massive, K-band selected, galaxies with significant 1.4 GHz radio detections, and thus not strictly IR-selected as it is assumed above although all 6 sources share similar IR luminosities. The actual density of such BzK sources is thus likely lower than the space density of IR sources in the field. Hence, this procedure should be taken with caution and only for comparison purposes.

Figure \ref{fig:lumfunction} shows the space density of CO sources as a function of the average CO $1-0$ luminosity obtained for our covered field as well as the space density of CO sources from the 6 BzK galaxies of \citet{Daddi2010a}, compared to the predictions from \citet{Obreschkow2009b} for $z=0$, $z=1$ and $z=2$ and observations of local galaxies by \citet{Keres2003}. Similar to the findings of \citet{Obreschkow2009b}, the space density of CO sources derived from the BzK galaxies is only marginally consistent with the one predicted by the simulations. Furthermore, the space density obtained for the CO sources in our field appears to be in excess from that predicted in the simulations and from that obtained for the BzK sample. This result comes naturally from the fact that our observations likely target an overdensity of galaxies at $z\sim1.5$. If we correct the $\sim7\times$ increase factor due to the overdense field, we obtain a value in agreement, within the uncertainties, with what is measured from the BzK galaxies and marginally consistent with expectations from simulations. Similarly, if we take into account only the two spectroscopically confirmed CO sources in our field, the space density and average CO luminosity tend to agree better with the estimate obtained from the BzK galaxies. The apparent offset between measurements and the predictions from simulations could be explained by several reasons or a combination of them, for instance: (a) lack of a numerous sample of CO sources; (b) although the Obreschkow et al. model assumes a galactic CO luminosity to gas mass conversion factor, the use of different values could produce a better agreement between predictions and observations; and (c) source-to-source variations in the line ratios, e.g. between $2-1$ and $1-0$. These results exemplify the need to perform measurements of CO emission, particularly CO $1-0$, in a significant sample of galaxies (ideally in a contiguous field) selected based on their IR luminosities, in order to constrain the evolution of the CO luminosity function and thereby models of galaxy formation and evolution.

\subsection{Future searches for blind CO line emission}

The determination of redshifts for galaxies based on (sub)millimeter and radio wavelength observations have been focused on the IR brightest sources, favored by the expanded bandwidth of such facilities. Based on the relationship between IR and CO luminosities, an IR bright galaxy, such as an SMG, is likely to also have bright CO emission. This facilitates a high significance ``blind''  CO emission line detection along the line of sight, leading to an accurate redshift measurement. In such studies, the position of their targets are known a priori, usually derived from large blank-field IR/submm surveys, permitting to spatially guide the CO observations, being thus not really ``blind'' CO detections \citep{Weiss2009, Lestrade2010,  Lupu2010, Frayer2011,Scott2011, Cox2011, Riechers2011b, Combes2012}. However, serendipitous CO line detections, where the locations and redshifts of the gas-rich galaxies were  not anticipated have been reported \citep{Daddi2009a, Daddi2009b, Lestrade2010}.

Blind spectroscopic imaging of the sky with large bandwidths have a huge potential for discovery of gas-rich, optically and IR faint galaxies, however they require deep integrations to ensure significant emission line detections, at the $\sim10\sigma$ level, and unequivocally identify the galaxies and their redshifts. In the absence of such high signal-to-noise detections, observations need to be guided by well-known optical, IR or radio positions and by accurate photometric redshifts. As shown in this work, focused line observations of clustered fields at high-redshift have the advantage to increase the likelihood of finding gas emission using a limited bandwidth. This approach reduces the required observing time compared to purely blank fields, where besides the need to cover a substantial area of the sky, it is necessary either to cover a reasonable range in frequency (e.g. a full 3 mm band for ALMA, a full K or Q band with JVLA) in order to contain emission lines at different redshifts, or to perform ``spot-checks'' at a given redshift.

\section{Summary and concluding remarks}

In this work, we have presented deep CO $1-0$ line observations of galaxies located in a galaxy cluster candidate at $z=1.5$ using the JVLA. The candidate cluster was identified using the red-sequence technique and is associated to an overdensity of $\sim\times7$ of star-forming galaxies. We use the spatial clustering and expanded bandwidth of the JVLA to simultaneously observe the CO emission from four galaxies with available optical spectroscopic redshifts in a single pointing and frequency setting. 

We detect, at the $\sim3\sigma$ and $\sim4\sigma$ level, the CO $1-0$ emission line from two of the galaxies with available spectroscopic redshifts: ID 51613 and ID 51858 at $z=1.516$ and $z=1.556$, respectively. We find $L'_\mathrm{CO}=(3.9\pm1.1)\times10^{10}$ K km s$^{-1}$ pc$^2$ and $L'_\mathrm{CO}=(1.3\pm0.4)\times10^{10}$ K km s$^{-1}$ pc$^2$, respectively. Both galaxies are also detected with {\it Spitzer} at 24 $\mu$m, and with {\it Herschel} at 250 $\mu$m and 350 $\mu$m. While modeling of their dust properties indicates total IR luminosities of $5.7\times10^{11}\ L_\odot$ and $9.5\times10^{11}\ L_\odot$, respectively, characterization of their multi-wavelength SEDs suggests both are young star-forming galaxies that have formed a significant fraction of their stellar content, with stellar masses of $4.5\times10^{10}\ M_\odot$ and $6.0\times10^{10}\ M_\odot$, respectively. 

We performed a blind search for significant ($>4\sigma$) emission line peaks in the JVLA data cube around the position of optical sources that were selected to have photometric redshift in the range $z=1.2-1.8$ and limited to $K_S<23.5$. This selection criteria ensures an accurate photometric redshift, while selecting galaxies with stellar masses $>2\times10^9\ M_\odot$. We find that 2 of these selected optical galaxies are associated with significant emission line peaks ($>4\sigma$), which are thus identified as CO $1-0$ line emission candidates. Both galaxies, ID 51207 and ID 51380, are found to have photometric redshifts in the range $1.4-1.6$. Based on Monte Carlo simulations of the distribution of such emission line peaks in the field, we find that any optical galaxy that follows our selection criteria and is located within $1''$ from a candidate emission line peak is statistically unlikely to be associated to such peak by chance. In the case of ID 51207, such probability is $3.7\%$ while for ID 51380 we find 10\%. Using in parallel a Bayesian inference approach to select emission line peaks in the data cube, we recover both CO-emitting candidate sources. Here, the line identified with ID 51380 is found to have a probability of $70\%$ of being a CO emission line, however the line identified with ID 51207 is found to have a rather low probability of 13\%.  

We find CO based spectroscopic redshifts of 1.530 and 1.551 with $L'_\mathrm{CO}=(1.03\pm0.25)\times10^{10}$ K km s$^{-1}$ pc$^2$ and $L'_\mathrm{CO}=(1.12\pm0.26)\times10^{10}$ K km s$^{-1}$ pc$^2$ for ID 51207 and ID 51380, respectively. These galaxies, ID 51207 and ID 51380, are detected at 24 $\mu$m and 250 $\mu$m with {\it Spitzer} and {\it Herschel}. Such measurements lead to IR luminosities of $4.9\times10^{11}\ L_\odot$ and $3.3\times10^{11}\ L_\odot$ for ID 51207 and ID 51380, respectively. In the former case, its optical/IR SED suggest a $\sim$1 Gyr old, very massive star forming galaxy, with a stellar mass of $1.1\times10^{11}\ M_\odot$, while for ID 51380 the SED indicates that it corresponds to a young star-forming galaxy, with a stellar mass of $2.8\times10^{10}\ M_\odot$. 

In one case, we associated a line peak candidate with an optical source that had a previous optical spectroscopic redshift. This secure optical redshift of $z=1.240$ is incompatible with the one that would be derived from the identification of the line peak with the CO $1-0$ emission line. This optical source fits well within our source selection criteria, $K<23.5$ (AB) and $z_\mathrm{phot}=1.2-1.8$, however if we had identified it with the CO line emission, the non-detection in the IR bands and its implied SFR from the SED fitting method would have implied a very low SFE. We argue that in the absence of high-significance line detections, tight restrictions should be imposed on sources to be identified with CO line emission, such as detection in the IR bands or a lower limit in SFRs. 

We measured the space density of CO galaxies compared to the space density of CO emitters estimated from the 6 BzK galaxies significantly detected in CO emission by Daddi et al., and compared to predictions from semi-analytic simulations. Overall, we find that all observations are only roughly consistent with the simulations, despite the low number of detections of typical star-forming galaxies at high-redshift. Clearly, observations of molecular gas from a statistically significant sample of these galaxies is necessary to measure the evolution of CO luminosity function with redshift, and thus constrain models of galaxy formation and evolution. We conclude by discussing the advantages of performing deep CO observations of star-forming galaxies in clustered fields compared to blank-fields in the sky. 

\section*{Acknowledgments}

We thank the anonymous referee for the comments and suggestions that helped to improve this manuscript. MA and this work was co-funded under the Marie Curie Actions of the European Commission (FP7-COFUND). CC thanks the Kavli Institute for Cosmology for their hospitality. Based on observations obtained, within the COSMOS Legacy Survey, the National Radio Astronomy Observatory Jansky Very Large Array (JVLA),  Canada-France-Hawaii Telescope (CFHT), Subaru, Kitt Peak National Observatory (KPNO), Cerro Tololo Inter-American Observatory (CTIO), and ESO Observatories. The National Radio Astronomy Observatory is a facility of the National Science Foundation (NSF), operated under cooperative agreement by Associated Universities Inc.

\bibliographystyle{mn2e}
\bibliography{cluster_evla}

\bsp

\label{lastpage}

\end{document}